# DRR4Covid: Learning Automated COVID-19 Infection Segmentation from Digitally Reconstructed Radiographs

Pengyi Zhang, Yunxin Zhong, Yulin Deng, Xiaoying Tang, Xiaoqiong Li

## Abstract

Automated infection measurement and COVID-19 diagnosis based on Chest X-ray (CXR) imaging is important for faster examination. However, due to the heterogeneous nature of CXRs and the difficulty of precisely annotating, most previous studies have developed classification models rather than segmentation models for COVID-19 diagnosis, and taken the advantages of the interpretability of classification model (e.g., saliency map) to locate the infected regions in lungs roughly. To address this problem, we propose a novel approach, called DRR4Covid, to learn automated COVID-19 diagnosis and infection segmentation on CXRs from digitally reconstructed radiographs (DRRs). Specifically, we design DRR4Covid with a modular framework, which comprises of an infection-aware DRR generator, a classification and/or segmentation network, and a domain adaptation module. The infection-aware DRR generator is able to produce DRRs with adjustable strength of radiological signs of COVID-19 infection, and generate pixel-level infection annotations that match the DRRs precisely, thus enabling the segmentation networks to be trained directly for automated infection segmentation. Although the synthetic DRRs are photo-realistic, there is still a gap between synthetic DRRs and real CXRs, which may lead to a poor model performance on real CXRs. The domain adaptation module is introduced to solve this problem by training networks on unlabeled real CXRs and labeled synthetic DRRs together. Due to the modular framework, our DRR4Covid can be assembled flexibly with off-the-shelf classification and segmentation networks, and domain adaptation algorithms. In this paper, we provide a simple but effective implementation of DRR4Covid by using a domain adaptation module based on Maximum Mean Discrepancy (MMD), and a FCN-based network with a classification header and a segmentation header. Extensive experiment results have confirmed the efficacy of our method; specifically, quantifying the performance by accuracy, AUC and F1-score, our network without using any annotations from CXRs has achieved a classification score of (0.954, 0.989, 0.953) and a segmentation score of (0.957, 0.981, 0.956) on a test set with 794 normal cases and 794 positive cases. Besides, we estimate the sensitive of X-ray images in detecting COVID-19 infection by adjusting the strength of radiological signs of COVID-19 infection in synthetic DRRs. The estimated detection limit of the proportion of infected voxels in the lungs is 19.43%±16.29%, and the estimated lower bound of the contribution rate of infected voxels is 20.0% for significant radiological signs of COVID-19 infection. Our codes will be made publicly available at https://github.com/PengyiZhang/DRR4Covid.

## 1. Introduction

The highly contagious Coronavirus Disease 2019 (COVID-19), caused by the severe acute respiratory syndrome coronavirus 2 (SARS-CoV-2) virus [1][2][3], has spread rapidly spread to most countries in the world. Globally, as of 2:34pm CEST, 7 July 2020, there have been 11,500,302 confirmed cases of COVID-19, including 535,759 deaths, reported to World Health Organization (WHO) [4]. Rapid detection and confirmation of COVID-19 infection is critical to prevent the spread of this epidemic.

The real-time reverse-transcriptase polymerase chain reaction (RT-PCR) is regarded as the golden standard for COVID-19 diagnosis [5][6]. However, the clinical findings has implied that RT-PCR has a low sensitivity [7][8][9], especially at the initial presentation of COVID-19.

Radiological imaging, such as Computed Tomography (CT) and Chest X-ray (CXR), is currently used to provide visual evidence for confirming positive COVID-19 patients in clinical. CT scan provides accurate 3D images of the lungs that are able to detect very small lesions effectively such as lung nodule and tumor. Existing CT findings in COVID-19 infection [10] have indicated that CT screening presents superior sensitivity over RT-PCR [8]. However, the workflow of CT imaging, involving several pre-scan events [11], is relatively complex and CT examinations are costly. As the number of infected patients rapidly increases, the routine use of CT brings heavy burden to the radiology department [12], which is not conductive to rapid screening of COVID-19. In comparison, CXR examination is much easier, faster and less costly, and provide high-resolution 2D images of the lungs that can detect a variety of lung conditions such as pneumonia, emphysema and cancer. Therefore, CXRs are the typical first-line imaging modality used for patients under investigation of COVID-19 [9], despite being less sensitive than initial RT-PCR testing (69% versus 91%, respectively) [13]. Developing automated infection measurement and COVID-19 diagnosis based on CXRs is important for faster examination.

Many approaches have been proposed for automated COVID-19 diagnosis based on CXRs, and have claimed notable detection accuracy of COVID-19 infection. However, the majority of these approaches are developed based on classification models rather than segmentation models, and some of these methods use saliency map (or attention map) to indicate the infected regions roughly. Therefore, these methods cannot provide precise segmentation of COVID-19 infection for further assessment and quantification. On the other hand, many approaches for infection segmentation have been developed based on CT imaging and some 3D CT scans with voxel-level infection annotations are publicly available. Compared with CT, infection segmentation based on CXRs is more challenging due to the heterogeneous nature of X-ray images and the difficulty of precisely annotating. Currently, there is no method developed for segmenting X-ray images for COVID-19 as reviewed by Shen et al. in [9].

Digitally reconstructed radiograph (DRR) [14][15][16][17] is a synthetic X-ray image that are generated by simulating the passage of X-rays through a 3D CT volume in specific poses (position and orientation) within a virtual imaging system. CXR findings on COVID-19 infection reflect the similar radiological signs with those described by CT [13] such as bilateral, peripheral consolidation and/or ground glass opacities (GGOs) [6][8][9]. Thus, we expect to take advantages of the public CT scans with voxel-level annotations of COVID-19 infection and the correlation between DRR and CXRs to realize automatic infection segmentation based on CXRs.

To this end, we propose a novel approach, called DRR4Covid, which can learn automated COVID-19 infection segmentation on CXRs from labeled DRRs. Specifically, we design DRR4Covid with a modular framework, comprising of an infection-aware DRR generator, deep classification and segmentation networks, and a domain adaptation module. Given a CT volume with voxel-level infection annotations, our infection-aware DRR generator is able to produce DRRs with adjustable level of COVID-19 infection, and simultaneously generate pixel-level infection annotations that match the corresponding DRRs accurately. Although such synthetic DRRs are photo-realistic, there is still a gap between synthetic DRRs and real CXRs, which may lead to a poor model performance on real CXRs. To improve the infection segmentation performance, we introduce the domain adaptation module to enable the networks to be trained on unlabeled real CXRs and labeled DRRs together. In this paper, we provide a simple but effective implementation of DRR4Covid by using a domain adaptation module based on Maximum Mean Discrepancy (MMD), and a FCN-based [18] network with a classification header and a segmentation header. Extensive experiment results have confirmed the efficacy of our methods; specifically, quantifying the performance by accuracy, AUC and F1-score, our network without using any annotations from CXRs has achieved a classification score of (0.954, 0.989, 0.953) and a segmentation score of (0.957, 0.981, 0.956) on a test set with 794 normal cases and 794 positive cases.

To our best knowledge, this is the first attempt to realize the automated COVID-19 infection segmentation base on CXRs by using the labeled DRRs that are generated from Chest CT scans. Due to the modular framework, *DRR4Covid* can be assembled flexibly with other off-the-shelf deep models, and domain adaptation algorithms. Moreover, *DRR4Covid* is a unified approach that can adapt to other lesion segmentation (e.g., lung nodule and tumor) based on X-ray imaging, called *DRR4Lesion*. Besides, we estimate the sensitive of X-ray images in detecting COVID-19 infection by adjusting the infection level of synthetic DRRs; we report the estimated detection limit of the infected proportion of the lungs is 19.43%±16.29%, and the estimated lower bound of the contribution rate of infected voxels is 20.0% for significant COVID-19 infection.

## 2. Related work

In this section, we review the related work from three aspects, including DRR, domain adaptation and CXR based screening of COVID-19 in a view of infection segmentation.

### 2.1 DRR

A digitally reconstructed radiograph (DRR) [14][15][16][17] is a synthetic X-ray image that are generated by simulating the passage of X-rays through a 3D CT volume in specific poses (position and orientation) within a virtual imaging system. DRRs are generally used as reference images by the intensity based two-dimensional to three-dimensional (2D–3D) image registration algorithms to verify the correct setup position of a patient for many radiotherapy treatments [19][20][21]. Each pixel value of DRR is obtained by calculating the radiological path length (RPL) [22], i.e., the summation of the length travelled by this ray in each voxel, multiplied by the relative CT intensity of the voxel that are measured in Hounsfield units (HUs). Thus, with a high complexity level of $O(n^3)$, DRR generation is computationally intensive by nature [23]. Meanwhile, in the iterative optimization of 2D–3D image registration algorithms, the generation process of DRR is usually performed many times to calculate the similarity measure [21], which greatly limits the running speed of 2D-3D image registration algorithms [17]. Therefore, the majority of researches have focused on this problem and proposed plenty of improved approaches to accelerate the generation of DRR [16][17][21]-[25]. In comparison, we are more concerned with the consistency between generated DRRs and their corresponding infection annotations. Thus, we design the infection-aware DRR generator of DRR4Covid directly based on SiddonGpuPy [26], which combines the serial algorithm proposed by Jacob [16] to improve the original Siddon's algorithm [14], and the parallel implementation proposed by Greef et al. [22].

The most closely related work is TD-GAN [27] and DeepDRR [28][29]. TD-GAN aims to learn automatic parsing of anatomical objects in X-ray images from labeled 3D CT scans by using synthetic labeled DRRs. The pixel-level labeling of DRR is generated by projecting 3D CT labels along the same trajectories used in relative DRR generation. TD-GAN adopts CycleGAN architecture to perform unpaired image-to-image translation and unsupervised domain adaptation, and thus the segmentation models trained on DRRs can generalize to real X-ray images. Similar strategy is also used by X2CT-GAN [30] to reduce the gap of synthetic DRRs and real X-ray images. Unlike TD-GAN, DeepDRR aims to generate more realistic radiographs and fluoroscopy from 3D CT scans, and enable machine learning models trained directly on DeepDRRs to generalize to clinical data without the need for domain adaptation. DeepDRR combines machine learning models for material decomposition and scatter estimation in 3D and 2D respectively with analytic models for projection, attenuation, and noise injection. DeepDRR has been applied to anatomical landmark detection in pelvic X-ray and to simulate X-rays of the femur during insertion of dexterous manipulators in orthopedic surgery. Clearly, both TD-GAN and DeepDRR care more about the anatomical structure with stronger radiological signs rather than lesion regions with weaker radiological signs. Besides, it is much easier to maintain the consistency between synthetic DRRs and their corresponding annotation masks of anatomical structures. In comparison, we aim at learning segmentation of COVID-19 infection in CXRs from CT volumes. Specifically, we design the infection-aware DRR generator that can enhance the weak radiological signs of COVID-19 infection, and maintain the consistency between generated DRRs and infection annotations through a category-weighted projection and RPL threshold method.

### 2.2 Domain Adaptation

Domain adaptation aims to rectify the distribution discrepancy between the training samples (source domain) and test samples (target domain) [31] and tune the model toward better generalization onto the target domain in a supervised or unsupervised manner. Numerous domain adaptation methods have been proposed for deep models recently as deep networks can learn more transferable features for domain adaptation and achieve better performance [32][33][34]. The main insight behind these approaches is to extract domain-invariant representations by embedding domain adaptation modules in the pipeline of deep learning [31][35][36][37][38][39][40][41].

Existing deep domain adaptation methods generally involve aligning the source and target domain distributions from three perspectives. The first stream is image alignment, and image-to-image translation models are typically used to reduce the gap between source domain images and target domain images [27]. The second stream is feature alignment [35][36][37][38][39][40], which is the majority approach and aims to learn domain-invariant deep features. The last stream is output alignment, which is often used to learn semantic segmentation of urban scenes from synthetic data [31][41]. On the other hand, we recognize that there are two main approaches to perform feature alignment, including adversarial approach [37][38][42][43][44] and non-adversarial approach [34][36][39][45][46][47]. The adversarial approach motivates deep models to extract domain-invariant features through adversarial training. It is done by training task-specific deep models to minimize the task-specific loss and the adversarial loss simultaneously, thereby fooling the domain discriminator to maximize the probability of deep features from source domain being classified as target domain. The non-adversarial approach is statistic moment matching-based approach, involving maximum mean discrepancy (MMD) [36][45][46], central moment discrepancy (CMD) [47] and second-order statistics matching [39]. The statistic moment matching-based approach encourages deep models to extract domain-invariant deep features by minimizing the distance between the statistic moments of deep features from source domain and from target domain. MMD [48] is the most representative method, and has been widely used to measure the discrepancy between the source and target distributions [34]. Compared with the adversarial approaches, MMD-based methods are simple, easy to implement, and thus facilitate to verify the efficacy of DRR4Covid quickly. In our implementation of DRR4Covid, we directly use an off-the-shelf MMD-based domain adaptation approach, i.e., LMMD proposed by Zhu et al [34], to enable the deep models trained on DRRs to generalize to real CXRs.

## 2.3 CXR based screening of COVID-19 in a view of infection segmentation

Segmentation is an essential step in automated infection measurement and COVID-19 diagnosis, which can provide the delineation of the regions of interest (ROIs), e.g., infected regions or lesions, in the CXRs for further assessment and quantification. Many approaches have been proposed for automated COVID-19 diagnosis based on CXRs. However, the majority of these approaches are developed based on classification models rather than segmentation models as reviewed by Shen et al. in [9] due to the aforementioned reasons. Thus, some researches resort to the interpretability of deep classification models to highlight the COVID-19 infection regions rather than accurately segmenting the infection regions. Specifically, Oh et al. [12] introduce a probabilistic Grad-CAM saliency map to highlight the multifocal lesions within CXRs in their local patch-based deep classification models for COVID-19 diagnosis. Such method is derived from a famous explanation technique, i.e., gradient weighted class activation map (Grad-CAM), and can effectively locate the radiological signs of COVID-19 infection, such as the multifocal ground-glass opacifications and consolidations. Similarly, Karim et al. [49] use a revised Grad-CAM, i.e., Grad-CAM++, and layer-wise relevance propagation (LRP) [50] in classifying CXRs as Normal, Pneumonia and COVID-19 to highlight the class-discriminating regions in CXRs. Moreover, Tabik et al. [51] adopt multiple explanation techniques, including occlusion [52], saliency [53], input X gradient [54], guided backpropagation [55], integrated gradients [56], and DeepLIFT [57], to investigate the interpretability of deep classification models and highlight the relevant infection regions of pneumonia and COVID-19 separately. To sum up, these approaches based on explanation techniques are mainly used for the inspection of deep model's decision, and are probably not suitable for further assessment and quantification. In comparison, our *DRR4Covid* is able to train deep segmentation models for precise infection segmentation directly without the need for the pixel-level infection annotations of real CXRs.

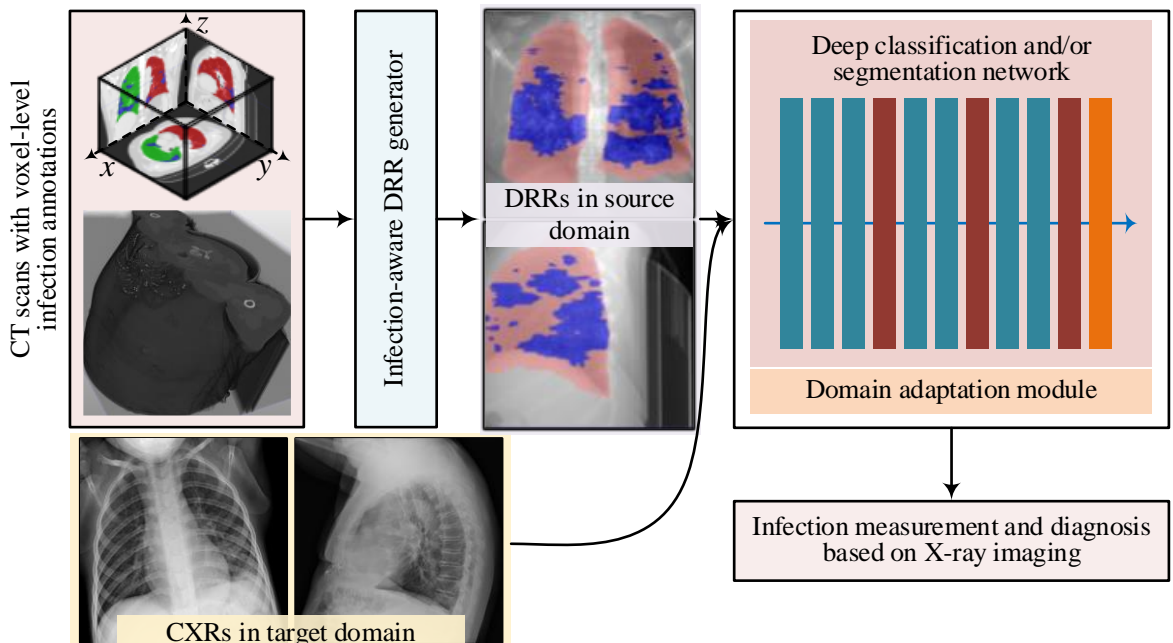

Figure 1. The modular framework of proposed DRR4Covid.

## 3. Methods

In this section, we describe the modular framework of proposed DRR4Covid, and analyze the key points in the design, followed by an introduction of our implementation of DRR4Covid to realize learning automated infection segmentation on CXRs.

### 3.1 Modular framework of DRR4Covid

Given CT scans with voxel-level infection annotations and unlabeled CXRs, we aim to learn deep models to perform automated COVID-19 infection segmentation on CXRs. We design DDR4Covid with a modular framework as shown in **Fig. 1**. DRR4Covid consists of three key components, including an infection-aware DRR generator, a deep model for classification and/or segmentation, and a domain adaptation module. The basic workflow of DDR4Covid involves generating DRRs with pixel-level infection annotations from CT scans, and training deep models on synthetic labeled DRRs and unlabeled CXRs by using the domain adaptation module.

**Generating labeled DRRs**. The DRR generator is responsible for synthesizing photo-realistic DRRs that resemble real CXRs as much as possible and producing pixel-level infection annotations that match the DRRs precisely by projecting 3D CT annotations along the same trajectories used in synthesizing DRRs. High-quality DRRs with pixel-level infection annotations in the context of this paper can be defined by two conditions. One is a good consistency between synthetic DDRs and their corresponding infection annotations; the other one is a good correlation between the visual findings of COVID-19 infection in synthetic DRRs and in real CXRs. As CXRs are generally considered less sensitive than 3D CT scans [9], it may happen that CT examination detects an abnormality, whereas the X-ray screening on the same patient reports no findings. DRRs also suffer from such problem, and it will cause the inconsistency between synthetic DRRs and the corresponding infection annotations. This is the first key point that need more attention to obtain high-quality DRRs in the design of DRR4Covid. The second key point is the correlation between the visual findings of COVID-19 infection in real CXRs and in synthetic DRRs. Note that the synthetic DRRs and infection annotations are used later to train deep models for classification and segmentation. Thus, a large gap between the visual findings of COVID-19 infection in real CXRs and in synthetic DRRs will cause deep models on DRRs fail to generalize to real CXRs even if the domain adaptation module is applied.

**Training deep models with the domain adaptation module**. Although synthetic DRRs are photo-realistic, there is still a gap between DRRs and real CXRs. Thus, we introduce the domain adaptation module into the framework of DRR4Covid. According to the quality of synthetic labeled DRRs, the problem of training deep models with the domain adaptation module will be further divided into two distinct kinds of problems. One is deep domain adaptation with fully supervised learning in source domain (i.e., synthetic DRRs) and unsupervised learning in target domain (i.e., real CXRs); whereas the other one is deep domain adaptation with weakly supervised learning in source domain and unsupervised learning in target domain. The premise of the first problem is the good consistency between synthetic DDRs and infection annotations. If such premise is not well satisfied, it will turn into the second problem due to the inaccurate synthetic infection annotations. Compared with the second problem, the first problem is well defined, and has been extensively studied. In this paper, we mainly focus on solving the first problem. Thus, we first implement a high-quality DRR generator, i.e., the infection-aware DRR generator.

### 3.2 Infection-aware DRR generator

We design an infection-aware DRR generator to produce high-quality DRRs defined in **section 3.1**. The standard DRR generator takes a CT volume or an infection annotation volume in a specific pose (position and orientation) as input and outputs a DRR or an infection mask. In comparison, our DRR generator takes both a CT volume and its infection annotation volume as input and produce a labeled DRR as illustrated in **Fig. 2**. A ray is casted from the X-ray source through labeled CT volumes to the center of each pixel of DRR. Each pixel value of DRR is obtained by calculating the class-weighted RPL [22], i.e., the class-weighted summation of the length travelled by this ray within each voxel, multiplied by the relative CT intensity of the voxel that are measured in HU. The calculation of the $d$-th pixel of DRR $p_d$ is formulated as:

$$p_d = \frac{\sum_{(i,j,k)\in\Omega_d} l_{(i,j,k)} * \rho_{(i,j,k)} * w_{(i,j,k)}}{\sum_{(i,j,k)\in\Omega_d} w_{(i,j,k)} / |\Omega_d|}, \quad (1)$$

where $\Omega_d$ is the 3D index set of the voxels in the X-ray direction, $|\Omega_d|$ is the number of voxels in $\Omega_d$, $l_{(i,j,k)}$ represents the normalized length travelled by the ray within the $(i,j,k)$-th voxel, $\rho_{(i,j,k)}$, and $w_{(i,j,k)}$ denote the CT value and the weight of the $(i,j,k)$-th voxel, respectively. The weight of the $(i,j,k)$-th voxel is defined as:

$$w_{(i,j,k)} = \begin{cases} w_2, \text{if } m_{(i,j,k)} = 2 \\ w_1, \text{elif } m_{(i,j,k)} = 1 \\ w_0, \text{otherwise} \end{cases} \bigg| \; m_{(i,j,k)} \in \{0,1,2\}, \quad (2)$$

where $m_{(i,j,k)} \in \{0,1,2\}$ is the category of the $(i,j,k)$-th voxel, 0, 1 and 2 represent the background, lungs and COVID-19 infection respectively. Note that the infection-aware DRR generator will produce standard DRRs when the weights of all categories are equal. On the other hand, the label of $p_d$, $m_d$, is computed as:

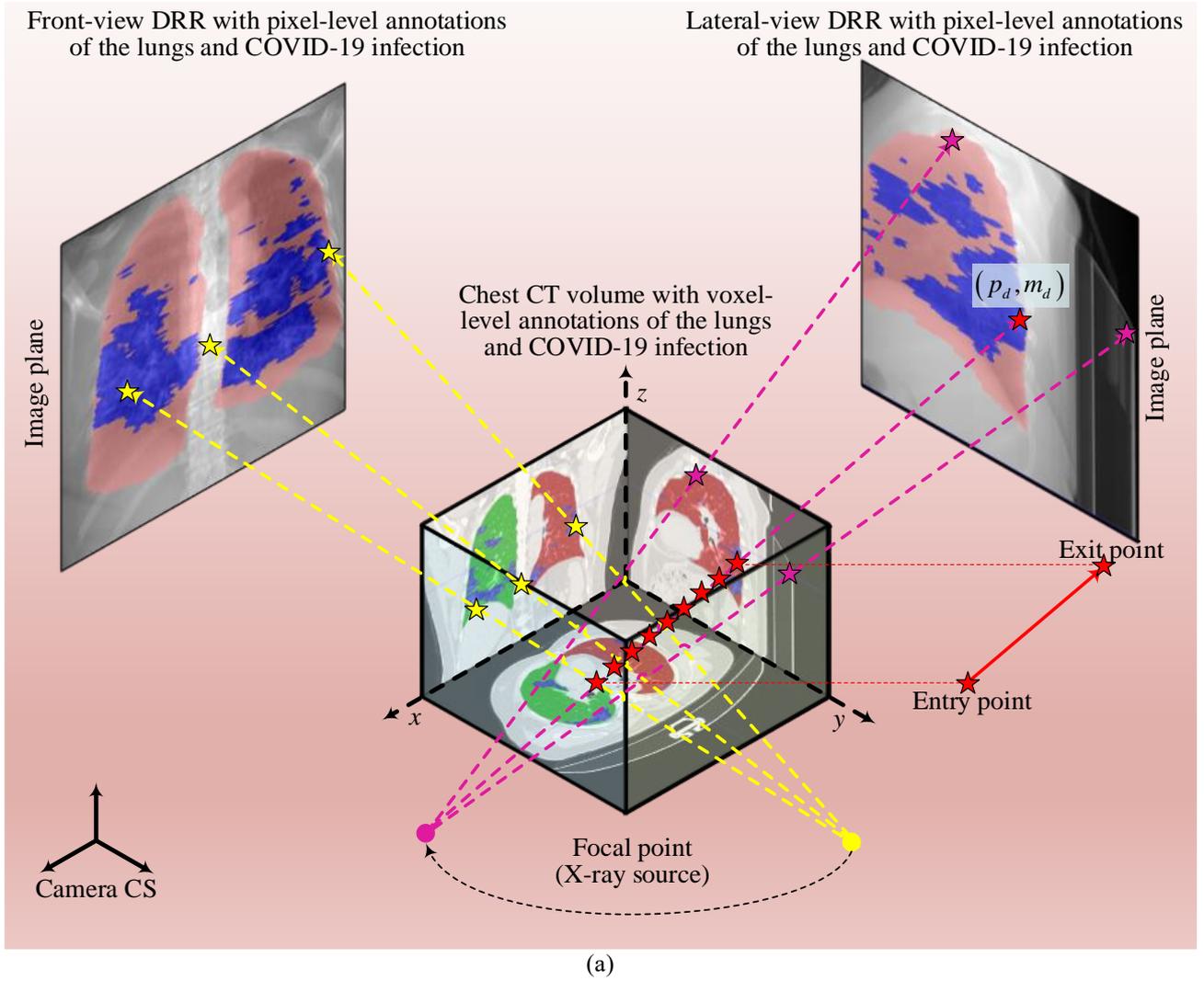

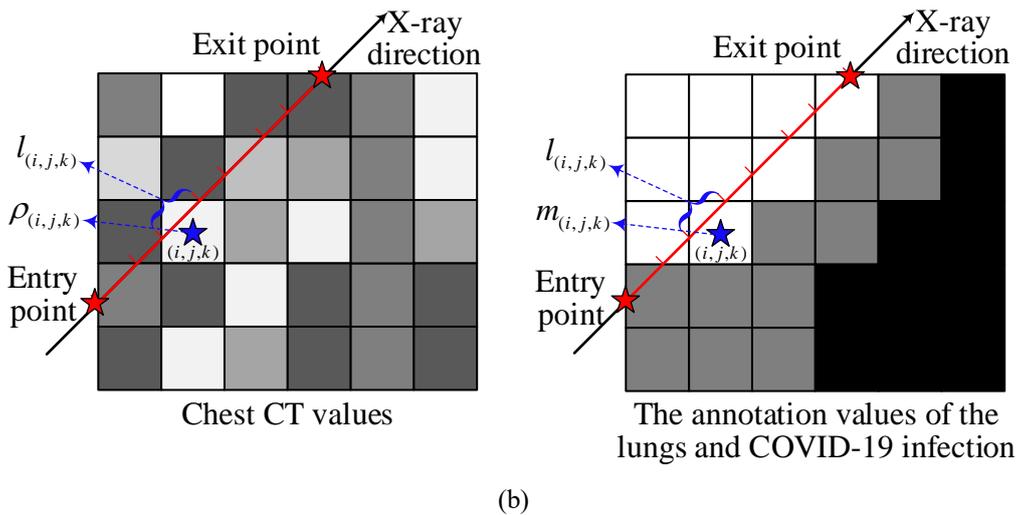

Figure 2. Illustration of the infection-aware DRR generation.

$$m_d = \begin{cases} 2, & \text{if } \pi_d^2 > T^2 \\ 1, & \text{elif } \pi_d^2 \leq T^2 \text{ and } \pi_d^1 > T^1 \\ 0, & \text{otherwise} \end{cases}, \quad (3)$$

where $\pi_d^c$ denotes the contribution rate of the voxels of category $c$ in calculating $p_d$, and $T^c$ represents the contrition threshold of category $c$. Specifically, $\pi_d^c$ is defined as:

$$\pi_d^c = \frac{\sum_{(i,j,k)\in \Omega_d^c} l_{(i,j,k)} * w_{(i,j,k)}}{\sum_{(i,j,k)\in \Omega_d} l_{(i,j,k)} * w_{(i,j,k)}} \Bigg| \, c \in \{0,1,2\}, \quad (4)$$

where $\Omega_d^c$ denotes the 3D index set of the voxels of category $c$ in the X-ray direction.

We argue that the strength of the radiological signs of COVID-19 infection in CXRs and DRRs depends on the contribution rate of infected voxels (CRIV) due to the heterogeneous nature of X-ray imaging. A higher value of CRIV means a larger number of infected voxels appear in the X-ray direction, and the radiological signs of COVID-19 infection, e.g., GGOs, become more significant. Such property of X-ray imaging is well modeled in formula (1) and (4) by our infection-aware DRR generator. Increasing the weight of infected voxels will improve the value of CRIV and vice versa. Accordingly, our infection-aware DRR generator can produce DDRs with different strengths of radiological signs of COVID-19 infection simply by adjusting the weight of infected voxel. Meanwhile, the synthetic pixel-level annotations of COVID-19 infection are also calculated based on the CRIV. Therefore, our infection-aware DRR generator can maintain the consistency between synthetic DRRs and annotations easily by using a large weight of infected voxels to improve the value of CRIV. Note that an excessive large value for the weight of infected voxels may lead to a large gap between the visual findings of COVID-19 infection in real CXRs and in synthetic DRRs so that even the domain adaptation module does not work.

To sum up, our infection-aware DRR generator has the following advantages:

(1) By setting the weight of infected voxels to a very small value, it can produce DRRs with no findings, which is essential for the training of deep classification models for COVID-19 diagnosis.

(2) By setting the weight of infected voxels to a relatively large value, it will generate high-quality DRRs with pixel-level annotations of COVID-19 infection, which is essential for the training of deep models for precise infection segmentation.

(3) By adjusting the weight of infected voxels from small values to large values, it will synthesize a serial of labeled DRRs with different strengths of the radiological signs of COVID-19 infection. Such DRRs might be able to be used to quantify the sensitivity of X-ray imaging in detecting COVID-19 infection.

### 3.3 FCN-based network equipped with a MMD-based domain adaptation module

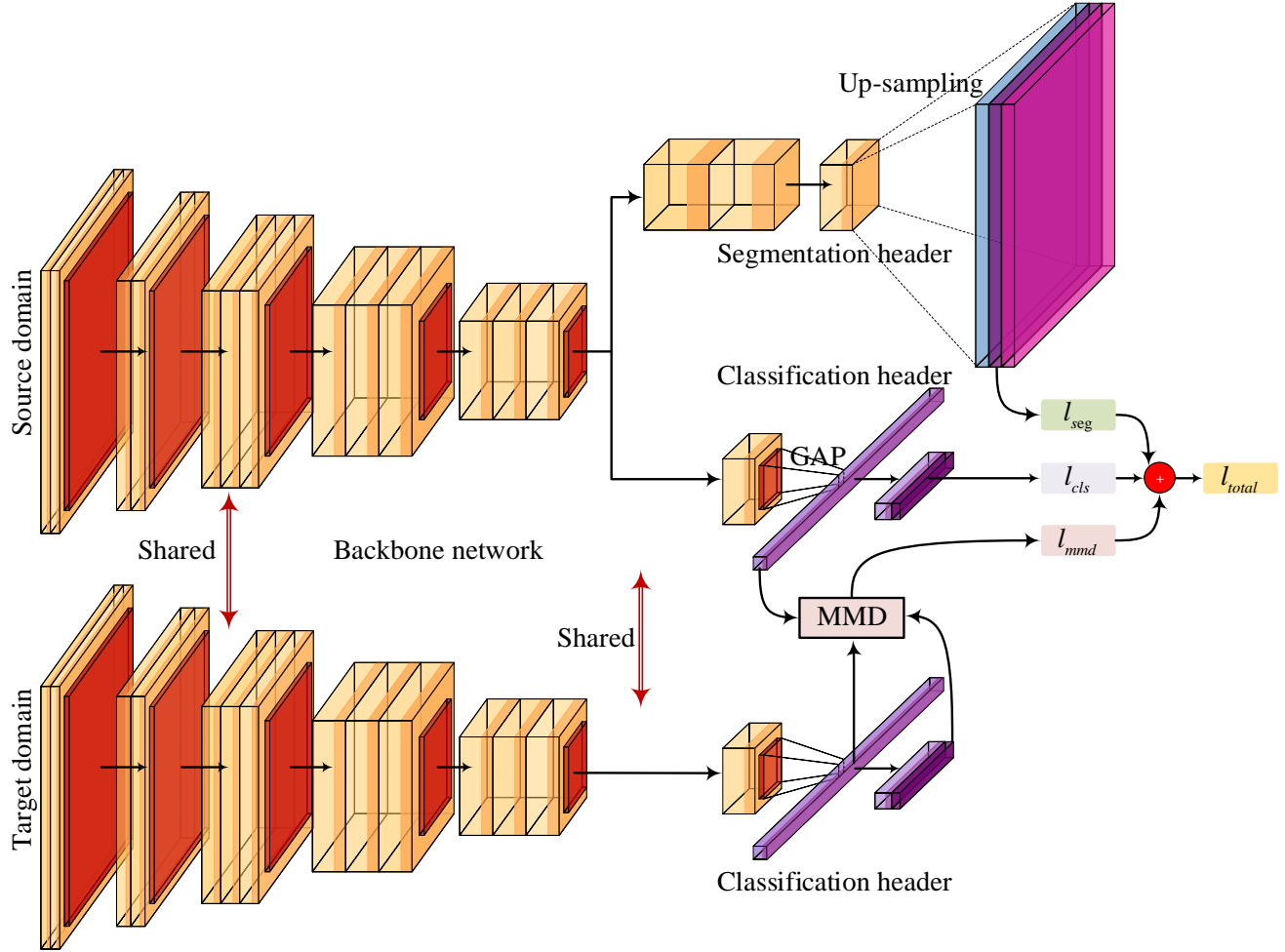

Figure 3. Illustration of the framework of our FCN-based network equipped with a MMD-based domain adaptation module.

**Network architectures**. We design a FCN-based network as depicted in **Fig. 3**. It consists of a backbone network, a classification header and a segmentation header. Compared with FCN [18], our model has an auxiliary classification header. The classification header is designed for two purposes. One is to enable our model to perform both classification task and segmentation task for automatic infection measurement and COVID-19 diagnosis. The other one is to facilitate the use of MMD-based methods for domain adaptation. The backbone network is responsible for extracting deep features by performing the convolution and spatial pooling operations on DRRs and CXRs. The extracted deep features are then fed into the classification header and segmentation header separately. In the classification branch, we adopt a very simple structure with a global average pooling (GAP) layer and a fully convolution (FC) layer. In the segmentation branch, we use two convolutional layers followed by an up-sampling layer to generate the segmentation output with the same size as the input DRRs and CXRs.

**MMD-based domain adaptation**. As a nonparametric distance estimate between two distributions, MMD [48] has been widely used in domain adaptation algorithms to measure the discrepancy between the source and target distributions. In our implementation, we adopt an off-the-shelf MMD-based domain adaptation approach, i.e., LMMD loss proposed by Zhu et al [34]. LMMD can measure the discrepancy of local distributions by taking the correlations of the relevant subdomains into consideration. By minimizing the LMMD loss in the training of deep models, the distributions of relevant subdomains within the same category in source domain and target domain are drawn close. As the LMMD method is proposed in the context of object recognition and digit classification tasks, we can apply it to the classification header directly by aligning the deep features from the GAP layer. The effect of feature alignment can be propagated to the segmentation branch implicitly through the input of the GAP layer.

**Objective function**. The training of our model is performed by minimizing the classification loss $l_{cls}$, segmentation loss $l_{seg}$ and LMMD loss $l_{mmd}$ simultaneously. The total loss is computed as:

$$l_{total} = \lambda_{cls} \times l_{cls} + \lambda_{seg} \times l_{seg} + \lambda_{mmd} \times l_{mmd} ,  \quad (5)$$

where $\lambda_{cls}$, $\lambda_{seg}$ and $\lambda_{mmd}$ denote the weights of the classification loss, segmentation loss and LMMD loss, respectively.

## 4. Experiments and Results

### 4.1 Materials

**Chest CT scans**. We use the public COVID-19-CT-Seg dataset [58], which consists of 20 public COVID-19 CT cases with pixel-level annotations of the left lung, right lung and COVID-19 infection. The annotations, first labeled by junior annotators, are refined by two radiologists with 5 years experience, and are further verified and refined by a senior radiologist with more than 10 years experience in chest radiology. In these 20 CT volumes, the voxel values of 10 volumes have been normalized to [0, 255] and we cannot access their CT values measured in HUs. We discard these ten cases and use the other 10 CT cases for DRR generation in our experiments. For each CT case, we are able to obtain 40 front-view DDRs and 40 lateral-view DDRs with pixel-level annotations of COVID-19 infection by using our infection-aware DRR generator, which will be detailed in **Section 4.2**. Thus, we build a training set in source domain with these 800 DRRs as shown in **Table 1**.

**Chest X-ray images**. We use two public CXR datasets, i.e., COVID-19 Radiography Database [59] and COVID-19 Chest X-ray Image Data Collection [60]. The former consists of 219 COVID-19 positive images and 1341 normal images, and the latter consists of 794 COVID-19 positive images. We randomly select 219 normal images from these 1341 normal images, and combine them with these 219 COVID-19 positive images in COVID-19 Radiography Database to build a train-validation set in target domain. Besides, we use the 794 COVID-19 positive images in COVID-19 Chest X-ray Image Data Collection and the 794 normal images that are randomly selected from the remaining 1122 normal images to build a test set in target domain as shown in **Table 1**. Moreover, 119 positive images and 119 negative (normal) images are randomly sampled from the train-validation set as training set, while the remaining images are used as validation set. All experiments are repeated five times with different splits of train-validation set, and the average results are reported.

**Table 1**. The split of training, validation and test sets.

| Source domain (DRRs) | Target domain (CXRs) | | |
|---|---|---|---|
| Training set | Training set | Validation set | Test set |
| 800 DRRs with pixel-level annotations of COVID-19 infection | 119 positive images 119 negative images | 100 positive images 100 negative images | 794 positive images 794 negative images |

### 4.2 Infection-aware DRRs

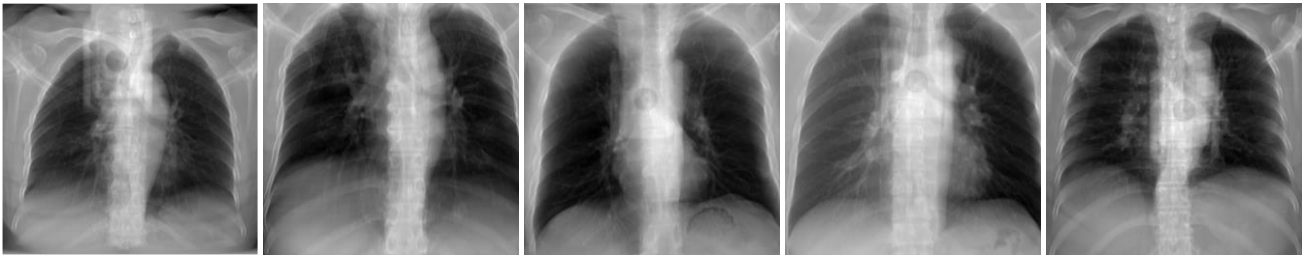

**Figure 4**. Illustration of normal DRRs generated by our infection-aware DRR generator.

**Generating normal DRRs.** DRRs with no findings are important for the training of deep classification and segmentation models for COVID-19 diagnosis. Our infection-aware generator is able to generate such DRRs with no findings by setting the weight of infected voxels to a relatively small value to reduce the CRIV in the ray-casting process. In our experiment, we empirically set the weights of background, lung and COVID-19 infection as $w_0=24.0, w_1=24.0, w_2=1.0$, and some synthetic normal DRRs are depicted in **Fig 4**.

**Generating multiple DRRs from each CT cases.** It is simple to generate multiple DRRs from a single CT volume by adjusting the pose (position and orientation) of the CT volume within a virtual imaging system. In our experiment, we randomly translate each CT volume between -100 and 100, and rotate it between -45° and 45° in *x*, *y*, and *z* directions. Some DRRs generated from a single CT case are illustrated in **Fig. 5**.

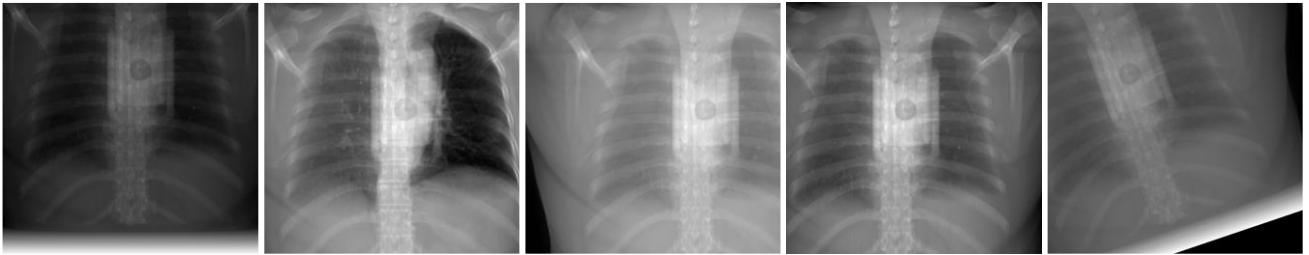

**Figure 5**. Illustration of multiple DRRs from a single CT case by adjusting the pose of the CT volume.

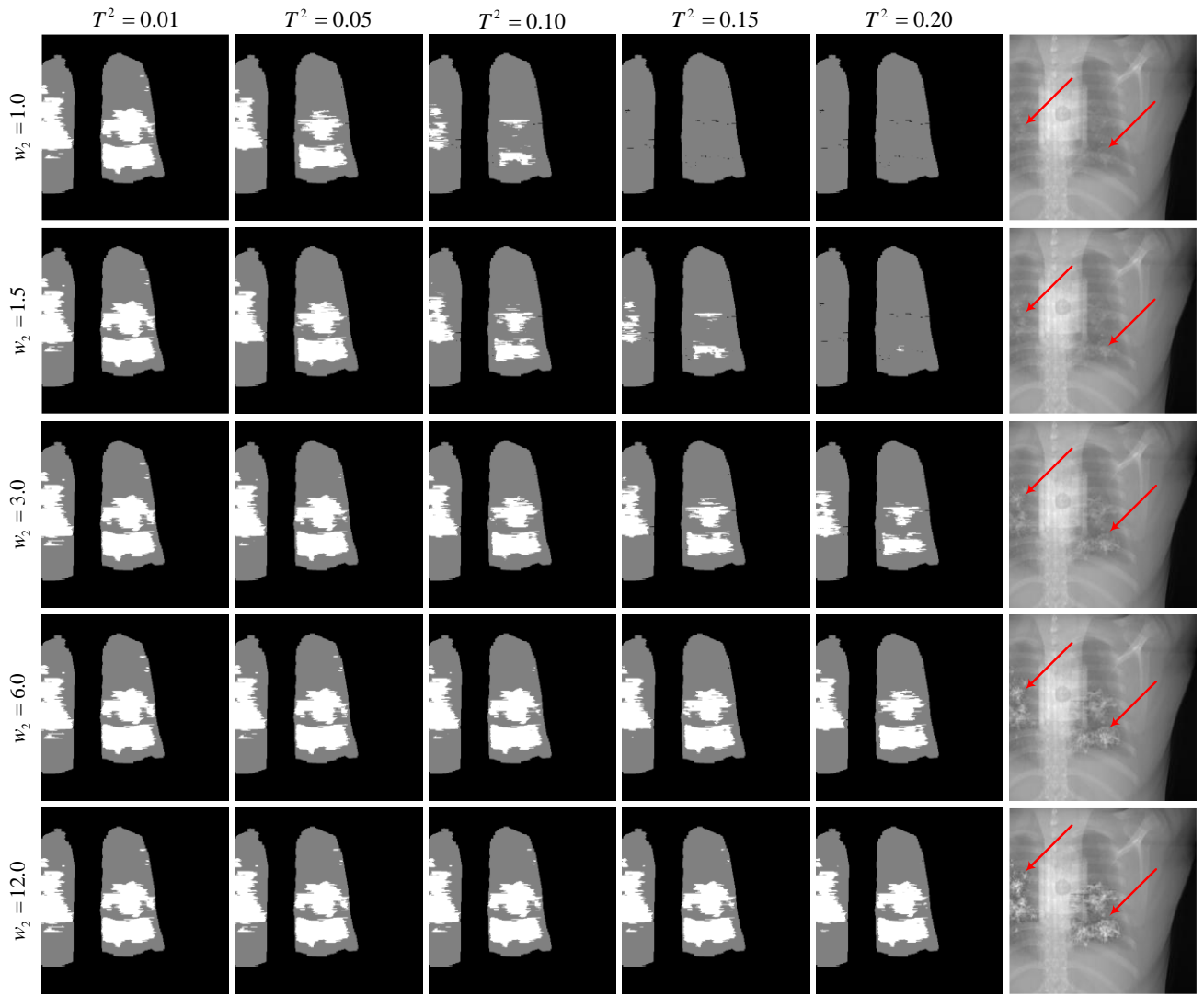

**Figure 6**. Illustration of DRRs with different strengths of radiological signs of COVID-19 infection from a single CT case and their corresponding pixel-level annotations of the lungs and COVID-19 infection. The weights of background and lungs are set as 1.0, and the contribution threshold of the lungs $T^1$ is set as 0.0. The red arrows in the last column highlight

the infected regions.

**Generating DRRs with different strengths of radiological signs of COVID-19 infection**. Our infection-aware DRR generator is able to generate DRRs with different strengths of radiological signs of COVID-19 infection by adjusting the weights of background, lung and COVID-19 infection $(w_0, w_1, w_2)$. In our experiment, we set $(w_0, w_1, w_2)$ to (12.0, 12.0, 1.0), (6.0, 6.0, 1.0), (3.0, 3.0, 1.0), (1.5, 1.5, 1.0), (1.0, 1.0, 1.0), (1.0, 1.0, 1.5), (1.0, 1.0, 3.0), (1.0, 1.0, 6.0) and (1.0, 1.0, 12.0) separately. Some samples are shown in the last column of **Fig 6**.

**Generating pixel-level annotations of COVID-19 infection**. We empirically set the contribution threshold of infected voxels (CTIV) $T^2$ as 0.00, 0.01, 0.05, 0.10, 0.15, 0.20 and 0.40 respectively to get the corresponding infection masks. The contribution threshold of the lungs is set to 0.00. Some infection masks are visualized in the first five columns of **Fig 6**.

**Building training sets in source domain (DRRs)**. For each CT volume, we first generate 40 normal DRRs, including 20 front-view DDRs and 20 lateral-view DDRs by randomly adjusting its pose. Next, in the same way we generate 40 DRRs with pixel-level annotations of COVID-19 infection with given $(w_0, w_1, w_2)$ and $T^2$. Therefore, we build a training set in source domain with 800 DRRs as shown in **Table 1**. Finally, with given the 63 different combinations of $(w_0, w_1, w_2)$ and $T^2$, we totally obtain 63 training sets in source domain.

### 4.3 Experiment setting

**Experiment design**. This paper aims at learning automated COVID-19 infection segmentation from DRRs. To this end, we propose DRR4Covid, which consists of an infection-aware DRR generator, a FCN-based network and a MMD-based domain adaptation module. To verify the efficacy of our method, we conduct experiments in three directions, including (1) standard DRRs versus infection-aware DRRs, (2) using of domain adaptation module versus no using of domain adaptation module, and (3) estimating the sensitivity of X-ray imaging in detecting COVID-19 infection by searching for the best parameters $(w_0, w_1, w_2)$ and $T^2$. Therefore, we first train the FCN-based network without using the MMD-based domain adaptation module separately on the 63 training sets in source domain respectively. Next, we train the same network with the MMD-based domain adaptation module separately on the 63 training sets in source domain and training set in target domain. All the trained models are finally evaluated on the same validation set and test set. We repeat all training tasks five times with different splits of train-validation set in target domain and report the average results. The tags of CXRs in target domain are kept unseen in all training tasks. Note that our infection-aware DRR generator will produce standard DRRs when $(w_0, w_1, w_2)$ equals to (1.0, 1.0, 1.0).

**Training details**. ResNet-18 is adopted as the backbone of the FCN-based network in our experiments. We train the network with 100 epochs by using Adam optimizer with the parameters of $\beta_1 = 0.9$ and $\beta_2 = 0.999$. We adopt mini-batch of 16, and use an initial learning rate of 0.0001 that is linearly decayed by 2% each epoch after 50 epochs. We initialize the backbone network with the weights of ResNet-18 pre-trained on ImageNet. Data augmentation, involving random cropping, horizontal flipping, vertical flipping and random rotating, are performed. The input image size of our network is 256×256×3. Besides, the category-weighted cross entropy loss is adopted to emphasize the optimization of COVID-19 infection segmentation, where the weights of background, lung and COVID-19 infection are set to 0.1, 1.0 and 5.0. The weights of the classification loss, segmentation loss and LMMD loss are set as $\lambda_{cls} = 1.0$, $\lambda_{seg} = 1.0$ and $\lambda_{mmd} = 0.3$, respectively.

**Evaluation metrics**. For the classification output of our model, we directly adopt the commonly used classification metrics, including accuracy, F1-score and area under precision-recall curve (AUC of PR-curve). As the pixel-level annotations of COVID-19 infection are not available in target domain (CXRs), we cannot use the segmentation evaluation metrics directly. To enable evaluate the quality of segmentation output of our model, we convert the segmentation output into classification output by determining whether there exists infected regions in the segmentation output, and then adopt the same three metrics.

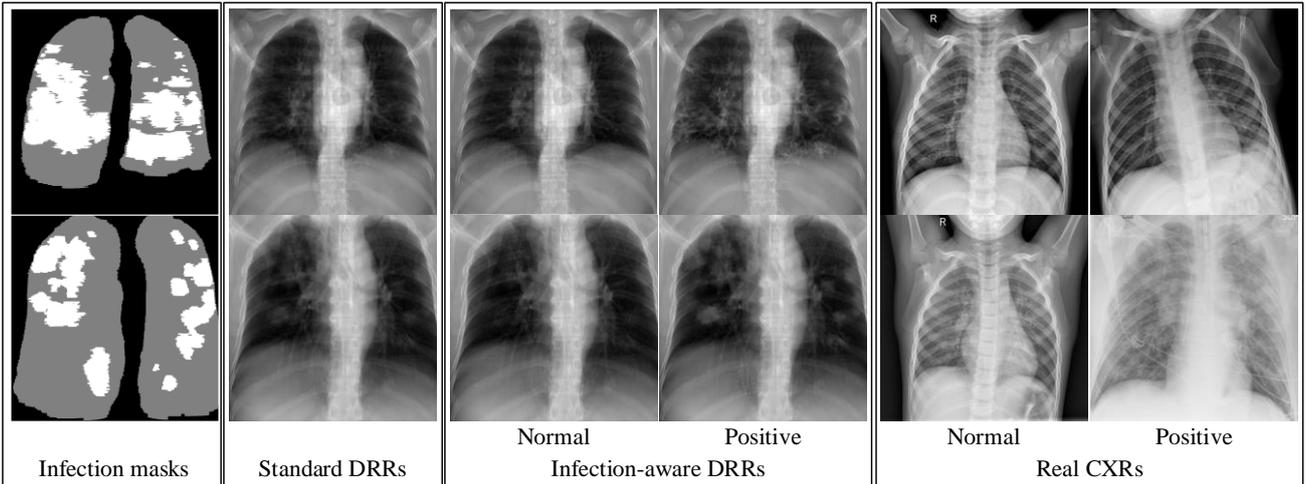

| Infection masks | Standard DRRs | Normal | Positive | Normal | Positive |
| | | Infection-aware DRRs | | Real CXRs | |

**Figure 7**. Comparison of standard DDRs, infection-aware DRRs and real CXRs. The first column represents the infection masks that are generated with the contribution threshold $T^2 = 0.00$. We use such infection masks to indicate the infected pixels of DRRs whose corresponding X-rays pass through the infected voxels of CT volume.

### 4.4 Experiment results

We report the evaluation results of our model trained on the 63 training sets with/without MMD-based domain adaptation module in **Table 1-24** of the appendix. We will analyze these results from the three perspectives as introduced in the experiment design.

**Standard DRRs versus infection-aware DRRs**. First, we do qualitative comparison in **Fig. 7**. As can be seen, many infected pixels of DRRs indicated by infection masks present no-findings due to the low contribution of infected voxels in the X-ray casting. This observation is consistent with the heterogeneous nature of X-ray imaging, and implicates the lower sensitivity of X-ray imaging in comparison of CT imaging. We notice that the radiological signs of COVID-19 infection in standard DRRs are rather weak. The strength of radiological signs of COVID-19 infection in standard DRRs depends only on the infected proportion of the lungs in CT volume. Such property makes it hard to take full advantages of the publicly available CT volumes. In comparison, our infection-aware DRR generator is able to produce DRRs with different strength of radiological signs of COVID-19 infection simply by adjusting the weight of infected voxels $w_2$. For instance, a CT case with mild COVID-19 infection can produce DRRs with strong radiological signs of COVID-19 infection, and a CT case with severe COVID-19 infection can get normal DRRs. Seen from the last column in **Fig. 6**, the radiological signs of COVID-19 infection get stronger gradually as the weight of infected voxels $w_2$ increases. Note that visual appearance of infected regions in DRRs will become unrealistic when the value of $w_2$ is too large, e.g., $w_2=12.0$, which may not facilitate to train deep models for automated COVID-19 infection segmentation. Such controllability of the strength of radiological signs of COVID-19 infection in DRRs is very helpful to make full use of the available CT volumes and to determine the precise infection masks to train infection segmentation models.

Next, we analyze the classification and segmentation results on validation and test sets without using domain adaptation in **Table 13-24** of the appendix. In order to avoid the interference of the choice of different contribution thresholds, we average the performance scores on CTIV, and compare the average scores of standard DRRs and infection-aware DRRs visually in **Fig. 8** and **Fig. 9**. The infection-aware DRRs achieve significantly higher average scores on both validation and test sets in target domain than the standard DRRs. Such results indicate that the gap between infection-aware DRRs (e.g., $w_2=3.0$) and real CXRs is smaller than the gap between standard DRRs ($w_2=1.0$) and real CXRs, and thus clearly verify the efficacy of our infection-aware DRR generator without using the domain adaptation module.

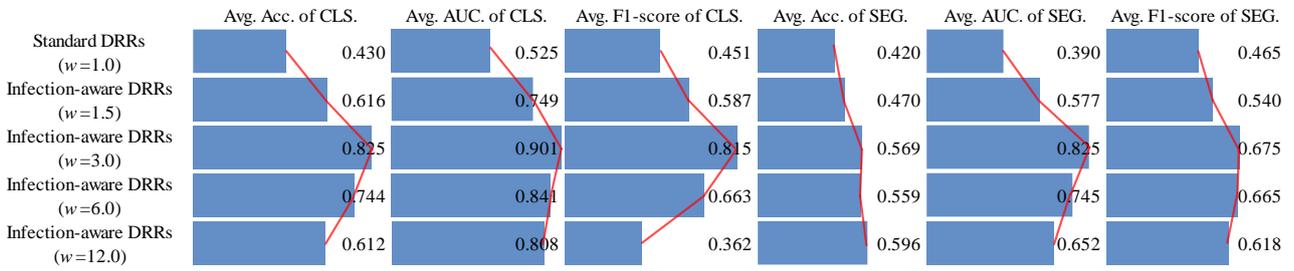

**Figure 8**. Comparison of average scores on validation set in target domain (no domain adaptation). CLS denotes classification results, SEG denotes segmentation results and $w$ represents the weight of infected voxels $w_2$. The scores are averaged on CTIV ($T^2=0.40, 0.20, 0.15, 0.1, 0.05, 0.01, 0.00$).

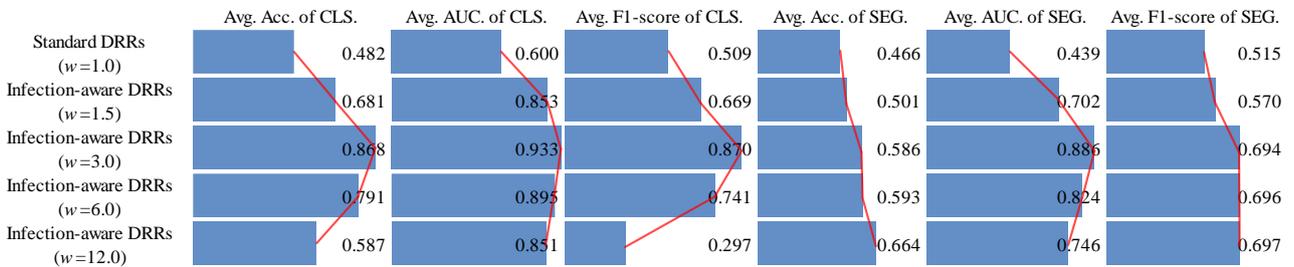

**Figure 9**. Comparison of average scores on test set in target domain (no domain adaptation). CLS denotes classification results, SEG denotes segmentation results and $w$ represents the weight of infected voxels $w_2$. The scores are averaged on CTIV ($T^2=0.40, 0.20, 0.15, 0.1, 0.05, 0.01, 0.00$).

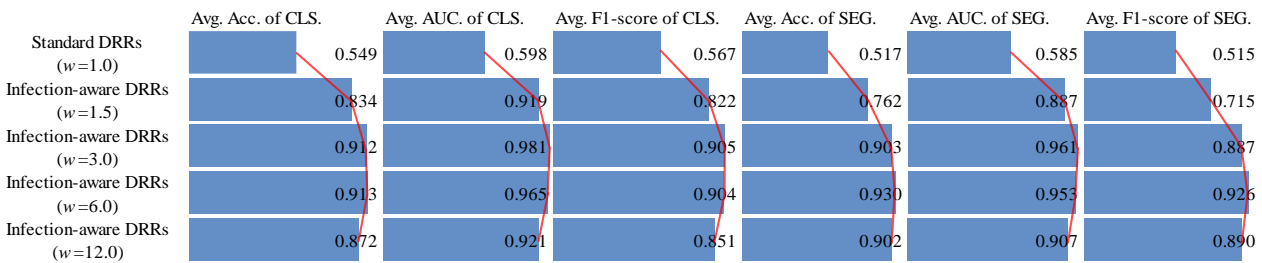

**Figure 10**. Comparison of average scores on validation set in target domain (domain adaptation). CLS denotes classification results, SEG denotes segmentation results and $w$ represents the weight of infected voxels $w_2$. The scores are averaged on CTIV ($T^2=0.40, 0.20, 0.15, 0.1, 0.05, 0.01, 0.00$).

Finally, we analyze the classification and segmentation results on validation and test sets with using domain adaptation in **Table 1-12** of the appendix. We compare the average results of standard DRRs and infection-aware DRRs visually in **Fig. 10** and **Fig. 11**. Similarly, the infection-aware DRRs surpass the standard DRRs by a large margin on both validation and test sets in target domain. Such results strongly demonstrates the effectiveness of our infection-aware DRR generator with using the domain adaptation module.

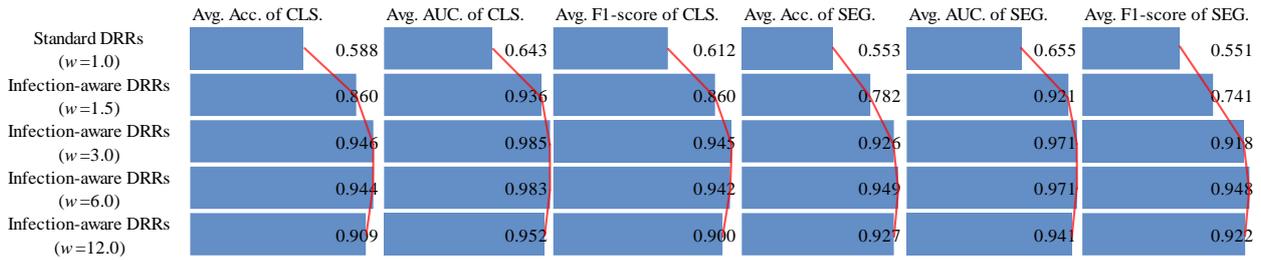

**Figure 11**. Comparison of average scores on test set in target domain (domain adaptation). CLS denotes classification results, SEG denotes segmentation results and $w$ represents the weight of infected voxels $w_2$. The scores are averaged on CTIV ($T^2 = 0.40, 0.20, 0.15, 0.1, 0.05, 0.01, 0.00$).

**Domain adaptation versus no domain adaptation**. In order to highlight the efficacy of our domain adaptation module, we compare the average scores of domain adaptation and no domain adaptation on validation and test sets in **Fig. 12** and **Fig. 13**. This intuitive comparison shows that the using of our domain adaptation module can improve the classification and segmentation scores of standard DRRs and infection-aware DRRs significantly and consistently, which strongly verifies the efficacy of our domain adaptation module. Besides, seen from **Fig. 8** and **Fig. 9**, we notice that the average scores of infection-aware DRRs increases first and then decrease as the weight of infected voxels $w_2$ increases from 1.0 to 3.0 and then to 12.0. The peak of average scores of infection-aware DRRs appears at $w_2 = 3.0$. It suggests that an excessively large weight of infected voxels may make the infected regions in DRRs unrealistic, thus leading to a decrease in performance scores without using domain adaptation module. In comparison, there is no significant decrease in the average scores of infection-aware DRRs with using our domain adaptation module as shown in **Fig. 10** and **Fig. 11** when the weight of infected voxels $w_2$ increases from 3.0 to 6.0 and then to 12.0. It implies that the domain adaptation module still works well even when infected regions in DRRs become slightly unrealistic. On the other hand, we observe that the segmentation scores are relatively lower than the classification scores without using the domain adaptation module. For instance, in the case of infection-aware DRRs with $w_2 = 3.0$, the average segmentation scores on the test set in target domain, including the accuracy, AUC and F1-score, are 0.586, 0.886 and 0.694 respectively, whereas the corresponding classification scores are 0.868, 0.933 and 0.870. Such results implicate that the segmentation header is much more sensitive to the domain discrepancy between DRRs and real CXRs than the classification header. By using the domain adaptation module, both the segmentation scores and classification scores are greatly improved; specifically, the improvement in segmentation scores is much more significant than the improvement in classification scores. For instance, in the case of infection-aware DRRs with $w_2 = 3.0$, the average segmentation scores on the test set are 0.926 (+0.340↑), 0.971 (+0.085↑) and 0.918 (+0.224↑) respectively, whereas the corresponding classification scores are 0.946 (+0.078↑), 0.985 (+0.052↑) and 0.945 (+0.075↑). Such results indicate that our domain adaptation module works well not only for classification task but also for segmentation task.

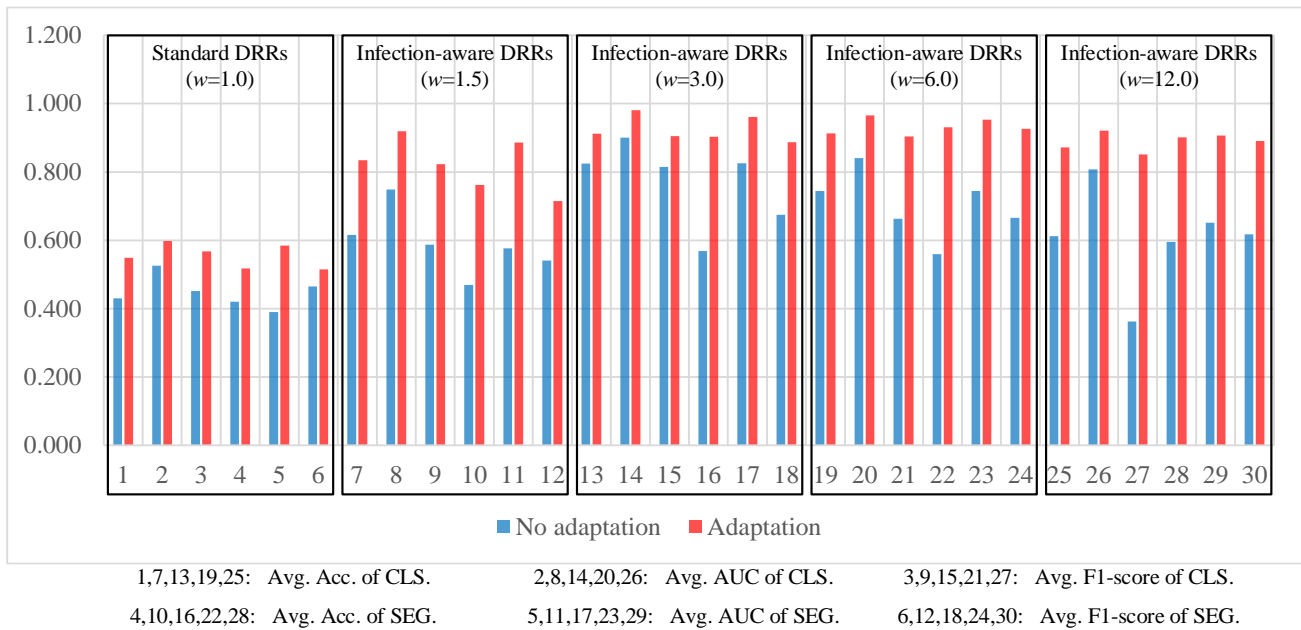

1,7,13,19,25: Avg. Acc. of CLS.  2,8,14,20,26: Avg. AUC of CLS.  3,9,15,21,27: Avg. F1-score of CLS.
4,10,16,22,28: Avg. Acc. of SEG.  5,11,17,23,29: Avg. AUC of SEG.  6,12,18,24,30: Avg. F1-score of SEG.

**Figure 12**. Comparison of average scores on validation set with domain adaptation and without domain adaptation. The scores are averaged on CTIV ($T^2 = 0.40, 0.20, 0.15, 0.1, 0.05, 0.01, 0.00$).

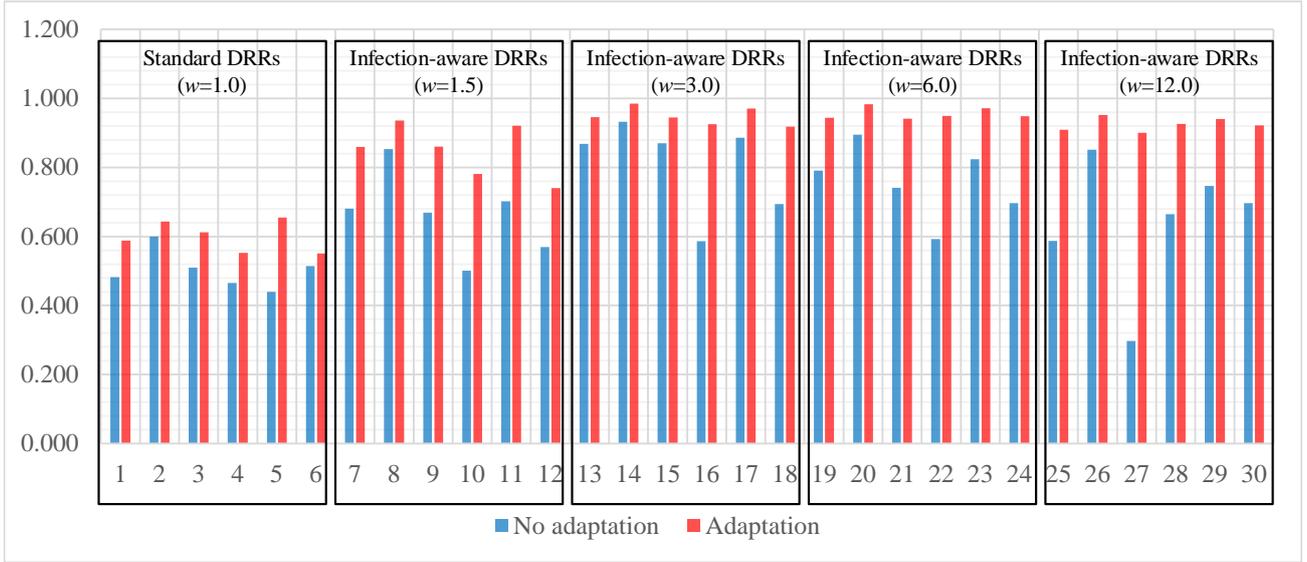

| | | |
|---|---|---|
| 1,7,13,19,25: Avg. Acc. of CLS. | 2,8,14,20,26: Avg. AUC of CLS. | 3,9,15,21,27: Avg. F1-score of CLS. |
| 4,10,16,22,28: Avg. Acc. of SEG. | 5,11,17,23,29: Avg. AUC of SEG. | 6,12,18,24,30: Avg. F1-score of SEG. |

**Figure 13**. Comparison of average scores on test set with domain adaptation and without domain adaptation. The scores are averaged on CTIV ($T^2 = 0.40, 0.20, 0.15, 0.1, 0.05, 0.01, 0.00$).

**Visualizing the COVID-19 infection segmentation results**. We specifically use the case of infection-aware DRRs with $w_2 = 3.0$ and $T^2 = 0.20$ as an example to show the visualized COVID-19 infection segmentation results. The corresponding segmentation scores, including accuracy, AUC and F1-score, on the validation and test sets are (0.945, 0.978, 0.943) and (0.957, 0.981, 0.956) respectively as listed in **Table 4**, **5**, **6**, **10**, **11** and **12** of the appendix. Note that these segmentation scores are averaged on five different training-validation splits. Next, we visualize the infection segmentation results of one of the five training-validation splits. The confusion matrices of the segmentation results on the corresponding validation and test sets are shown in **Fig. 14**. We visualize several true positive and true negative cases in **Fig. 15**. Compared with previous studies that highlight the infected regions roughly by resorting to the interpretability of deep classification models, our segmentation model trained on the infection-aware DRRs is able to segment the infected regions in CXRs directly and accurately. Besides, we present several failure (false positive and false negative) cases in **Fig. 16**.

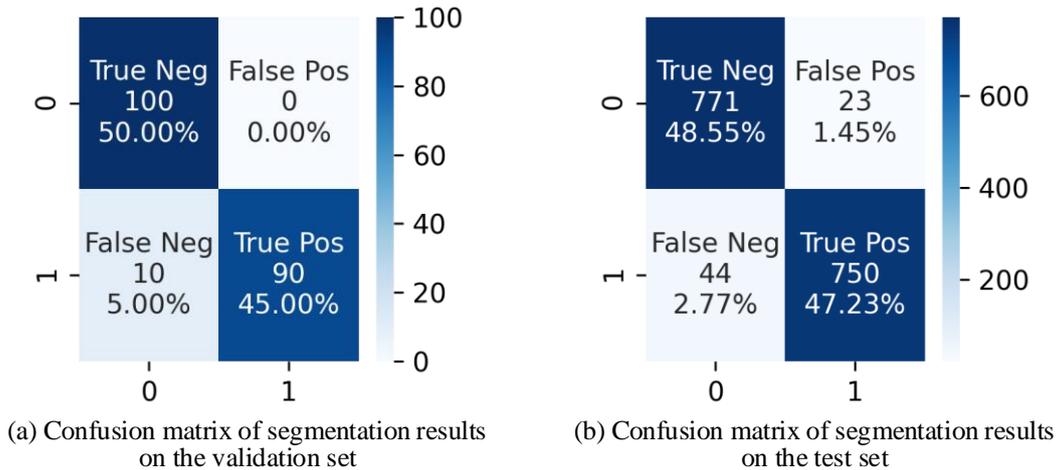

(a) Confusion matrix of segmentation results on the validation set

(b) Confusion matrix of segmentation results on the test set

**Figure 14**. Confusion matrices of segmentation results on the validation and test sets in the case of infection-aware DRRs with $w_2 = 3.0$ and $T^2 = 0.20$.

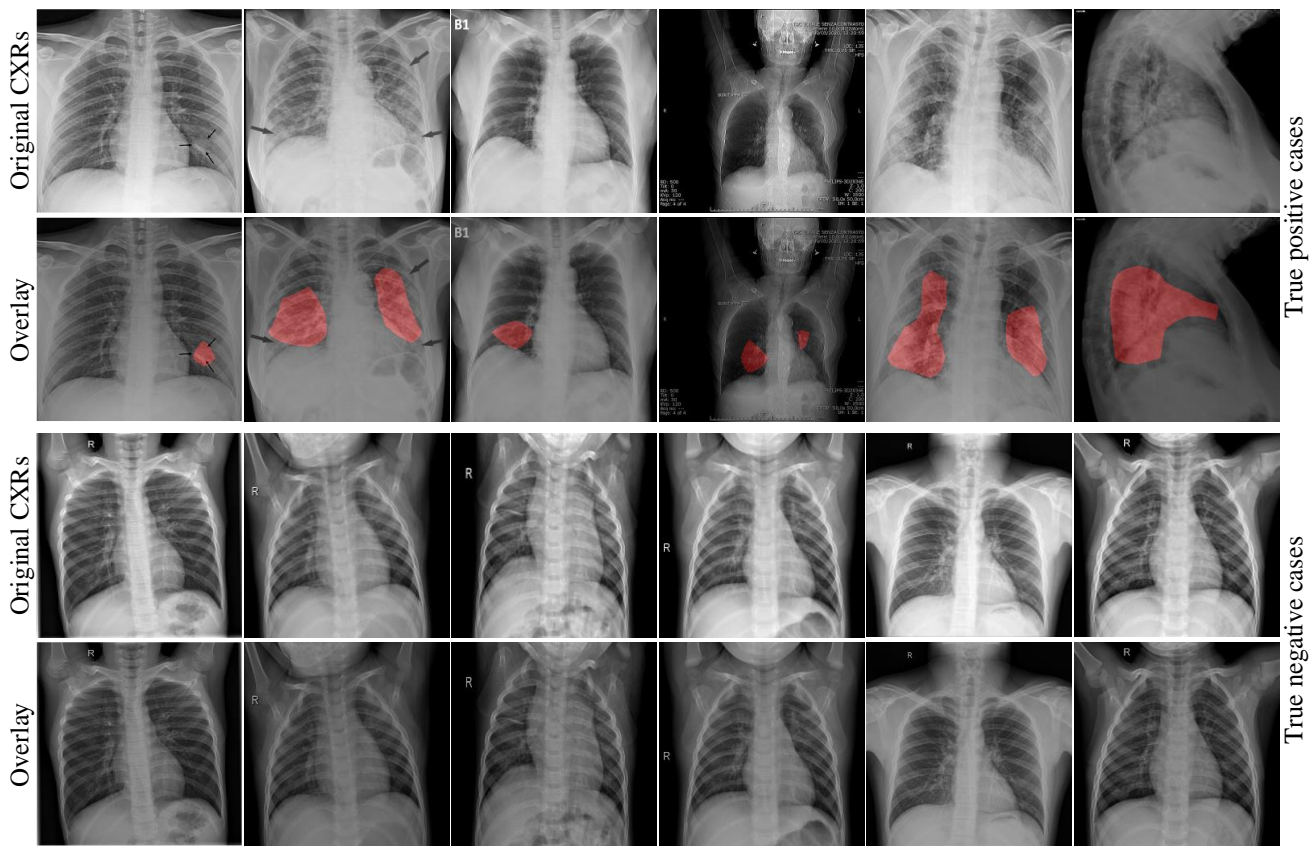

(a) Segmentation results on the validation set in target domain

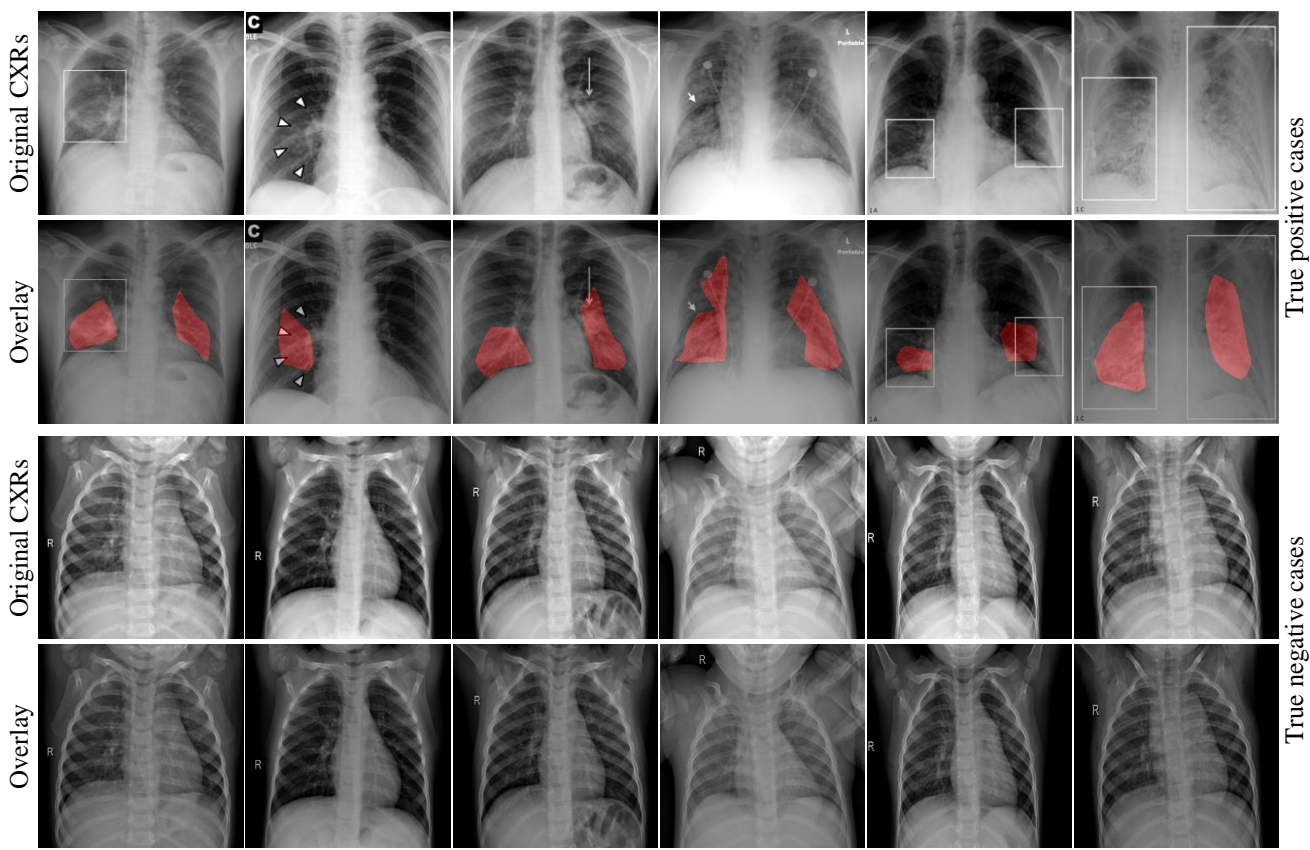

(b) Segmentation results on the test set in target domain

**Figure 15**. COVID-19 infection segmentation results of the infection-aware DRRs with $w_2 = 3.0$ and $T^2 = 0.20$ on the validation and test sets in target domain. The red overlay is used to indicate the infected regions.

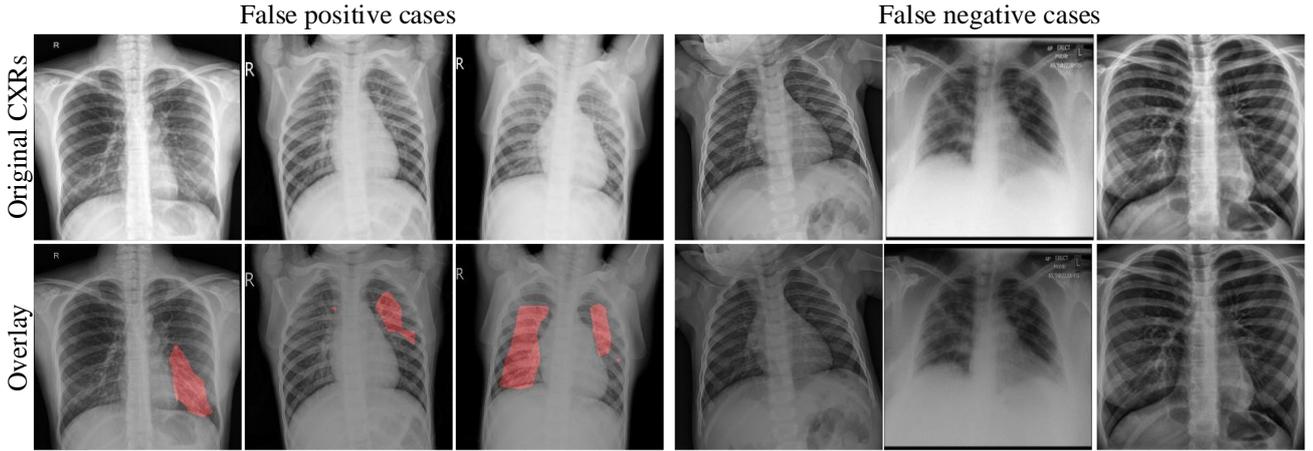

**Figure 16**. Failure cases of the segmentation results of the infection-aware DRRs with $w_2 = 3.0$ and $T^2 = 0.20$ on the validation and test sets. The red overlay is used to indicate the infected regions.

**Estimating the sensitivity of X-ray imaging in detecting COVID-19 infection by searching for the best parameters** $(w_0, w_1, w_2)$ **and** $T^2$. As mentioned earlier, CXRs are generally considered less sensitive than 3D CT scans [9]. It may happen that CT examination detects an abnormality, whereas the X-ray screening on the same patient reports no findings. To solve this problem, we introduce the infection-aware DRR generator to generate the infection-aware DRRs by increasing the weight of infected voxels to control the strength of radiological signs of COVID-19 infection in DRRs, and thus maintain the consistency of COVID-19 infection between generated DRRs and CT scans. We argue that the best parameters $(w_0, w_1, w_2)$ and $T^2$ will produce the best DRRs and get the highest classification and segmentation scores on the validation and test sets no matter whether the domain adaptation module is used or not. Therefore, we average the corresponding items in **Table 1-24** of the appendix to obtain the total average (T-Avg.) scores of 63 training sets in the source domain to search for the best parameters. Meanwhile, we compute the equivalent average proportion of infected voxels (EAPIV) in the lungs of the 10 CT cases that are used to generate infection-aware DRRs as shown in the last row of **Table 2**. As can be seen, the peak of T-Avg. scores at each row in **Table 2** appears consistently at $w_2 = 3.0$, and the corresponding EAPIV is 19.43%±16.29%. Such results indicate that the representative value of the proportion of infected lung is 0.194. It suggests that X-ray imaging can detect COVID-19 infection effectively when the proportion of infected voxels in the lungs of CT volume arrives at 19.43%±16.29%. Moreover, we plot the histograms of the CRIV of the pixels in infection-aware DRRs in **Fig. 17** to examine the function of the CTIV further. These histograms show the effectiveness of our infection-aware DRR generator in changing the distribution of CRIV of the pixels in infection-aware DRRs. For the best parameters $w_{0,1} = 1.0$ and $w_2 = 3.0$ (the 8-th column of **Table 2**), the peak of T-Avg. scores appears at $T^2 = 0.20$. Therefore, we argue that the lower bound of CRIV is 20.0% for significant radiological signs of COVID-19 infection in DRRs. It means the pixels whose CRIVs are lower than 20.0% can only present insignificant radiological signs of COVID-19 infection, and it is hard to distinguish such pixels from the pixels of the lungs.

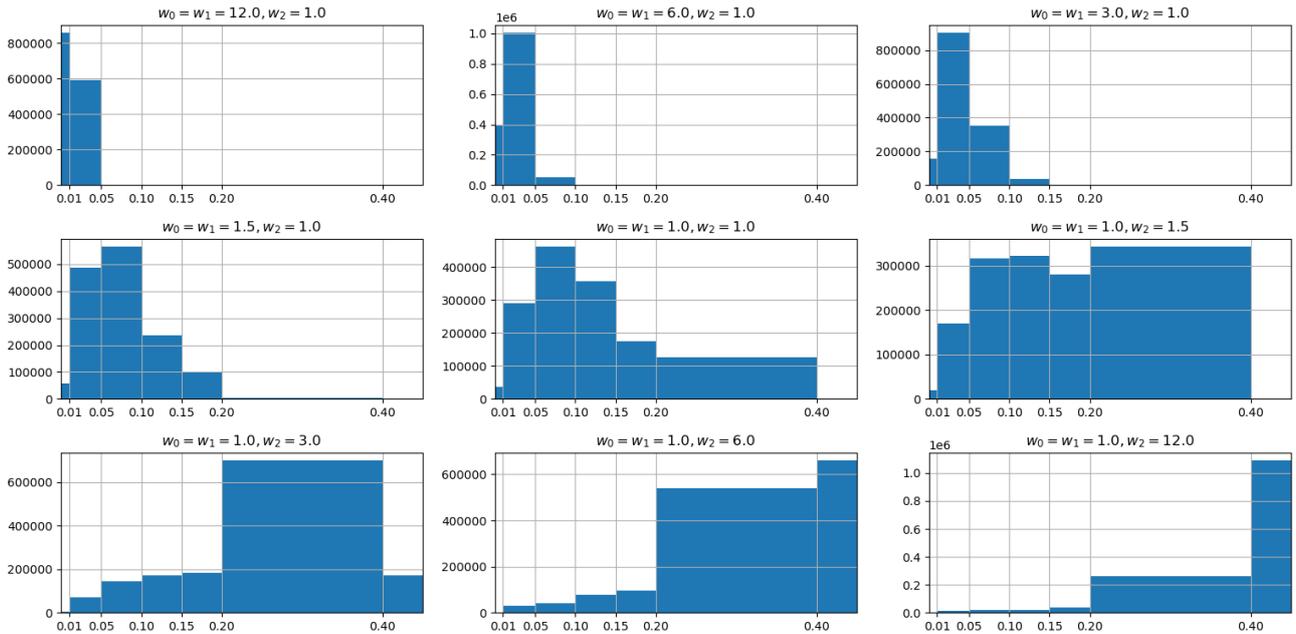

**Figure 17**. Histograms of the contribution rates of infected voxels of the pixels in infection-aware DRRs.

Table 2. Total average (T-Avg.) scores (Mean ± Standard Deviation) of our model trained on 63 different training sets in source domain. Each item in this table is the average of the corresponding items in **Table 1-24** of the appendix. EAPIV denotes the equivalent average proportion of infected voxels in the lungs of the 10 CT cases that are used to generate infection-aware DRRs.

| T-Avg. (%) | $w_{0,1}=12.0$ $w_2=1.0$ | $w_{0,1}=6.0$ $w_2=1.0$ | $w_{0,1}=3.0$ $w_2=1.0$ | $w_{0,1}=1.5$ $w_2=1.0$ | $w_{0,1}=1.0$ $w_2=1.0$ | $w_{0,1}=1.0$ $w_2=1.5$ | $w_{0,1}=1.0$ $w_2=3.0$ | $w_{0,1}=1.0$ $w_2=6.0$ | $w_{0,1}=1.0$ $w_2=12.0$ |
|---|---|---|---|---|---|---|---|---|---|
| $T^2=0.40$ | 29.5±22.2 | 31.2±23.2 | 31.4±23.7 | 35.0±25.8 | 43.2±31.4 | 54.7±30.9 | 79.5±12.5 | 83.7±14.4 | 76.1±18.7 |
| $T^2=0.20$ | 31.2±23.3 | 31.7±24.5 | 40.1±29.0 | 43.0±14.6 | 50.3±7.5 | 67.6±11.9 | **88.4±10.9** | 84.5±12.6 | 76.3±19.0 |
| $T^2=0.15$ | 32.8±23.9 | 36.7±27.1 | 32.4±21.7 | 42.2±6.7 | 48.0±8.4 | 76.4±14.5 | 87.2±12.4 | 83.0±12.6 | 77.4±18.6 |
| $T^2=0.10$ | 28.5±22.4 | 37.4±27.9 | 42.9±13.4 | 35.3±9.3 | 50.6±9.3 | 77.3±15.4 | 87.3±12.8 | 84.0±13.3 | 76.4±19.8 |
| $T^2=0.05$ | 40.7±31.6 | 48.1±20.9 | 27.7±9.4 | 33.8±11.4 | 57.5±10.8 | 77.7±13.3 | 87.4±11.9 | 84.2±12.7 | 77.9±17.5 |
| $T^2=0.01$ | 39.7±16.2 | 41.5±12.6 | 31.8±12.2 | 38.4±10.9 | 57.7±12.8 | 79.4±13.8 | 86.7±12.2 | 83.0±13.6 | 76.7±19.0 |
| $T^2=0.00$ | 44.6±14.9 | 42.4±12.6 | 32.2±12.4 | 40.6±11.0 | 61.0±14.0 | 78.9±15.5 | 86.8±12.2 | 83.4±12.8 | 75.5±17.5 |
| EAPIV (%) | 0.85±0.98 | 1.67±1.90 | 3.22±3.57 | 6.03±6.37 | 8.51±8.63 | 11.77±11.30 | 19.43±16.29 | 29.75±20.56 | 42.30±22.57 |

## Conclusion

We propose a novel approach, called DRR4Covid, to learn automated COVID-19 infection segmentation on CXRs from DRRs. DRR4Covid consists of three key components, including an infection-aware DRR generator, a classification and segmentation network, and a domain adaptation module. The infection-aware DRR generator is able to produce DRRs with adjustable strength of radiological signs of COVID-19 infection, and generate pixel-level infection annotations that match the DRRs precisely, thus enabling the segmentation networks to be trained directly for automated infection segmentation. The domain adaptation module is introduced to reduce the domain discrepancy between DRRs and CXRs by training networks on unlabeled real CXRs and labeled DRRs together. We provide a simple but effective implementation of DRR4Covid by using a domain adaptation module based on Maximum Mean Discrepancy (MMD), and a FCN-based network with a classification header and a segmentation header. Extensive experiment results have confirmed the efficacy of our methods; specifically, quantifying the performance by accuracy, AUC and F1-score, our network without using any annotations from CXRs has achieved a classification score of (0.954, 0.989, 0.953) and a segmentation score of (0.957, 0.981, 0.956) on a test set with 794 normal cases and 794 positive cases. Besides, we estimate the sensitive of X-ray images in detecting COVID-19 infection by adjusting the strength of radiological signs of COVID-19 infection in synthetic DRRs; we report that the estimated detection limit of the infected proportion of the lungs is 19.43%±16.29%, and the estimated lower bound of the contribution rate of infected voxels is 20.0% for significant COVID-19 infection.

To our best knowledge, this is the first attempt to realize the automated COVID-19 infection segmentation base on CXRs by using the labeled DRRs that are generated from Chest CT scans. The limitation of our work is that the segmentation results can only be evaluated by using classification metrics due to the unavailability of pixel-level annotations of COVID-19 infection in CXRs. Future work can be carried out by extending the DRR4Covid to DRR4Lesion to enable multiple lung lesion segmentation on CXRs.

## Appendix.

**Table 1**. Accuracy (Mean ± Standard Deviation) table of classification output on validation set in target domain (domain adaptation).

| Accuracy (%) | $w_{0,1}=12.0$ $w_2=1.0$ | $w_{0,1}=6.0$ $w_2=1.0$ | $w_{0,1}=3.0$ $w_2=1.0$ | $w_{0,1}=1.5$ $w_2=1.0$ | $w_{0,1}=1.0$ $w_2=1.0$ | $w_{0,1}=1.0$ $w_2=1.5$ | $w_{0,1}=1.0$ $w_2=3.0$ | $w_{0,1}=1.0$ $w_2=6.0$ | $w_{0,1}=1.0$ $w_2=12.0$ |
|---|---|---|---|---|---|---|---|---|---|
| $T^2=0.40$ | 50.0±0.0 | 50.0±0.0 | 50.0±0.0 | 50.0±0.3 | 52.6±2.9 | 69.2±5.6 | 89.6±1.2 | 94.5±1.3 | 89.7±2.1 |
| $T^2=0.20$ | 50.0±0.0 | 50.0±0.0 | 50.4±0.4 | 45.9±25.1 | 44.4±11.4 | 74.8±15.7 | 92.6±1.0 | 91.0±3.1 | 88.0±4.3 |
| $T^2=0.15$ | 50.0±0.0 | 50.0±0.0 | 40.3±17.8 | 39.4±17.4 | 46.4±16.4 | 86.8±8.2 | 90.8±1.4 | 89.8±5.0 | 87.0±4.8 |
| $T^2=0.10$ | 50.0±0.0 | 50.0±0.0 | 51.2±15.7 | 29.8±7.4 | 45.7±5.6 | 89.3±5.3 | 92.6±1.6 | 91.5±3.3 | 87.5±4.0 |
| $T^2=0.05$ | 50.0±0.0 | 56.7±1.3 | 23.6±3.3 | 24.6±2.7 | 61.1±14.8 | 85.6±7.6 | 92.1±2.2 | 91.3±4.2 | 87.4±4.2 |
| $T^2=0.01$ | 41.4±8.6 | 31.4±3.9 | 26.9±2.2 | 34.7±9.1 | 65.6±11.5 | 88.6±2.1 | 90.3±1.7 | 90.6±3.4 | 85.9±2.4 |
| $T^2=0.00$ | 49.2±10.2 | 35.9±2.4 | 28.4±5.1 | 40.1±11.5 | 68.3±7.6 | 89.6±2.9 | 90.3±1.5 | 90.1±2.5 | 84.8±5.1 |

**Table 2**. AUC (Mean ± Standard Deviation) table of classification output on validation set in target domain (domain adaptation).

| AUC (%) | $w_{0,1}=12.0$ $w_2=1.0$ | $w_{0,1}=6.0$ $w_2=1.0$ | $w_{0,1}=3.0$ $w_2=1.0$ | $w_{0,1}=1.5$ $w_2=1.0$ | $w_{0,1}=1.0$ $w_2=1.0$ | $w_{0,1}=1.0$ $w_2=1.5$ | $w_{0,1}=1.0$ $w_2=3.0$ | $w_{0,1}=1.0$ $w_2=6.0$ | $w_{0,1}=1.0$ $w_2=12.0$ |
|---|---|---|---|---|---|---|---|---|---|
| $T^2=0.40$ | 50.2±6.6 | 56.8±10.3 | 62.7±6.8 | 44.0±19.0 | 89.4±5.4 | 95.9±2.5 | 97.2±1.0 | 97.9±0.5 | 94.8±3.5 |
| $T^2=0.20$ | 58.1±8.9 | 70.0±8.2 | 69.1±13.1 | 46.8±39.0 | 40.5±20.1 | 75.1±29.8 | 97.7±1.0 | 96.2±2.0 | 92.2±6.1 |
| $T^2=0.15$ | 54.9±10.3 | 53.6±7.5 | 34.2±23.0 | 35.8±25.2 | 42.5±20.3 | 91.9±6.9 | 98.6±0.6 | 95.4±4.1 | 90.8±5.1 |
| $T^2=0.10$ | 49.2±8.7 | 74.2±14.5 | 52.0±24.5 | 17.6±7.2 | 33.8±6.7 | 95.4±3.2 | 98.8±0.6 | 97.5±1.8 | 91.9±7.4 |
| $T^2=0.05$ | 86.6±3.2 | 84.5±17.1 | 9.0±3.3 | 14.5±3.2 | 63.2±20.0 | 93.3±5.0 | 98.8±0.6 | 95.7±2.0 | 92.3±5.7 |
| $T^2=0.01$ | 18.3±9.5 | 19.7±5.4 | 12.1±3.6 | 26.0±8.4 | 69.6±17.0 | 94.8±1.8 | 97.6±2.4 | 96.4±1.6 | 90.7±4.3 |
| $T^2=0.00$ | 47.6±17.0 | 22.7±9.0 | 11.6±5.4 | 31.3±21.2 | 79.3±4.7 | 96.8±2.0 | 98.0±0.9 | 96.5±1.0 | 91.7±2.9 |

**Table 3**. F1-score (Mean ± Standard Deviation) table of classification output on validation set in target domain (domain adaptation).

| F1-score (%) | $w_{0,1}=12.0$ $w_2=1.0$ | $w_{0,1}=6.0$ $w_2=1.0$ | $w_{0,1}=3.0$ $w_2=1.0$ | $w_{0,1}=1.5$ $w_2=1.0$ | $w_{0,1}=1.0$ $w_2=1.0$ | $w_{0,1}=1.0$ $w_2=1.5$ | $w_{0,1}=1.0$ $w_2=3.0$ | $w_{0,1}=1.0$ $w_2=6.0$ | $w_{0,1}=1.0$ $w_2=12.0$ |
|---|---|---|---|---|---|---|---|---|---|
| $T^2=0.40$ | 0.0±0.0 | 0.0±0.0 | 0.0±0.0 | 0.4±0.8 | 9.6±10.3 | 54.6±11.4 | 88.5±1.6 | 94.2±1.4 | 88.6±2.7 |
| $T^2=0.20$ | 0.0±0.0 | 0.0±0.0 | 2.3±2.7 | 41.1±20.2 | 57.0±6.8 | 80.1±9.5 | 92.1±1.2 | 90.1±3.9 | 86.3±5.8 |
| $T^2=0.15$ | 0.0±0.0 | 0.0±0.0 | 11.6±6.4 | 48.7±14.7 | 57.3±10.5 | 87.6±6.1 | 90.0±1.7 | 88.5±6.6 | 84.8±6.5 |
| $T^2=0.10$ | 0.0±0.0 | 0.0±0.0 | 35.4±12.1 | 44.2±9.3 | 60.8±4.5 | 89.5±3.9 | 92.1±1.9 | 90.7±3.8 | 85.7±5.5 |
| $T^2=0.05$ | 0.0±0.0 | 23.8±3.7 | 35.9±4.9 | 37.8±3.5 | 69.0±8.5 | 86.7±5.8 | 91.5±2.5 | 90.4±5.3 | 85.3±5.5 |
| $T^2=0.01$ | 12.7±6.5 | 43.1±4.4 | 40.3±3.0 | 47.3±8.7 | 71.0±6.0 | 88.2±1.6 | 89.7±1.8 | 89.6±4.3 | 83.5±3.2 |
| $T^2=0.00$ | 26.7±3.3 | 47.7±0.7 | 41.9±6.3 | 52.9±8.5 | 72.4±4.6 | 89.0±3.0 | 89.7±2.0 | 89.1±3.1 | 81.8±6.7 |

**Table 4.** Accuracy (Mean ± Standard Deviation) table of segmentation output on validation set in target domain (domain adaptation).

| Accuracy (%) | $w_{0,1}=12.0$ $w_2=1.0$ | $w_{0,1}=6.0$ $w_2=1.0$ | $w_{0,1}=3.0$ $w_2=1.0$ | $w_{0,1}=1.5$ $w_2=1.0$ | $w_{0,1}=1.0$ $w_2=1.0$ | $w_{0,1}=1.0$ $w_2=1.5$ | $w_{0,1}=1.0$ $w_2=3.0$ | $w_{0,1}=1.0$ $w_2=6.0$ | $w_{0,1}=1.0$ $w_2=12.0$ |
|---|---|---|---|---|---|---|---|---|---|
| $T^2=0.40$ | 50.0±0.0 | 50.0±0.0 | 50.0±0.0 | 50.0±0.0 | 50.3±0.2 | 50.6±0.5 | 73.9±3.8 | 94.8±1.0 | 91.7±1.9 |
| $T^2=0.20$ | 50.0±0.0 | 50.0±0.0 | 50.0±0.0 | 44.8±8.8 | 48.9±8.5 | 66.9±14.7 | 94.5±0.7 | 92.8±2.6 | 90.9±3.3 |
| $T^2=0.15$ | 50.0±0.0 | 50.0±0.0 | 50.3±0.7 | 45.6±12.9 | 46.1±13.4 | 82.0±12.4 | 93.0±0.9 | 92.3±3.7 | 90.3±4.5 |
| $T^2=0.10$ | 50.0±0.0 | 50.1±0.2 | 51.5±4.1 | 27.0±4.6 | 44.9±3.4 | 83.7±9.3 | 93.2±1.5 | 94.5±1.5 | 90.9±2.9 |
| $T^2=0.05$ | 50.0±0.0 | 49.9±8.2 | 22.8±2.8 | 29.1±1.7 | 54.5±8.9 | 78.7±11.6 | 93.2±2.4 | 93.3±1.7 | 89.9±3.5 |
| $T^2=0.01$ | 40.3±1.3 | 41.1±2.1 | 35.0±1.8 | 38.5±4.8 | 58.0±8.4 | 84.3±5.8 | 92.9±1.5 | 92.2±3.2 | 89.0±2.8 |
| $T^2=0.00$ | 48.6±1.1 | 42.9±1.5 | 35.3±3.4 | 41.1±5.1 | 59.5±6.1 | 87.4±4.7 | 91.3±0.8 | 91.4±3.1 | 88.4±3.6 |

**Table 5.** AUC (Mean ± Standard Deviation) table of segmentation output on validation set in target domain (domain adaptation).

| AUC (%) | $w_{0,1}=12.0$ $w_2=1.0$ | $w_{0,1}=6.0$ $w_2=1.0$ | $w_{0,1}=3.0$ $w_2=1.0$ | $w_{0,1}=1.5$ $w_2=1.0$ | $w_{0,1}=1.0$ $w_2=1.0$ | $w_{0,1}=1.0$ $w_2=1.5$ | $w_{0,1}=1.0$ $w_2=3.0$ | $w_{0,1}=1.0$ $w_2=6.0$ | $w_{0,1}=1.0$ $w_2=12.0$ |
|---|---|---|---|---|---|---|---|---|---|
| $T^2=0.40$ | 33.1±15.2 | 37.9±16.9 | 45.6±11.3 | 47.4±10.7 | 77.3±10.4 | 87.7±5.4 | 90.6±1.7 | 96.8±1.3 | 93.0±2.6 |
| $T^2=0.20$ | 45.0±17.2 | 33.5±18.3 | 68.1±9.1 | 39.7±26.4 | 45.6±11.3 | 74.7±14.7 | 97.8±1.3 | 95.4±2.2 | 91.8±4.5 |
| $T^2=0.15$ | 49.3±17.8 | 60.2±10.8 | 41.7±17.6 | 43.3±17.0 | 45.4±15.6 | 90.0±7.9 | 96.9±1.4 | 93.7±6.7 | 90.4±5.6 |
| $T^2=0.10$ | 45.5±11.5 | 66.4±15.5 | 39.7±11.4 | 24.8±7.9 | 44.6±4.0 | 92.4±4.4 | 97.3±0.9 | 96.2±1.9 | 91.2±3.8 |
| $T^2=0.05$ | 75.3±6.7 | 50.8±15.9 | 13.3±4.5 | 18.8±4.9 | 63.2±11.5 | 90.9±5.8 | 97.6±0.2 | 95.7±2.5 | 90.8±5.5 |
| $T^2=0.01$ | 30.7±4.5 | 17.9±6.1 | 13.4±5.6 | 24.9±5.9 | 64.3±13.3 | 91.1±1.5 | 95.9±1.9 | 95.4±2.0 | 90.2±6.0 |
| $T^2=0.00$ | 28.7±6.9 | 16.7±6.3 | 14.1±7.1 | 26.7±18.9 | 69.1±5.5 | 93.8±0.5 | 96.5±1.0 | 93.7±6.6 | 87.4±7.1 |

**Table 6.** F1-score (Mean ± Standard Deviation) table of segmentation output on validation set in target domain (domain adaptation).

| F1-score (%) | $w_{0,1}=12.0$ $w_2=1.0$ | $w_{0,1}=6.0$ $w_2=1.0$ | $w_{0,1}=3.0$ $w_2=1.0$ | $w_{0,1}=1.5$ $w_2=1.0$ | $w_{0,1}=1.0$ $w_2=1.0$ | $w_{0,1}=1.0$ $w_2=1.5$ | $w_{0,1}=1.0$ $w_2=3.0$ | $w_{0,1}=1.0$ $w_2=6.0$ | $w_{0,1}=1.0$ $w_2=12.0$ |
|---|---|---|---|---|---|---|---|---|---|
| $T^2=0.40$ | 0.0±0.0 | 0.0±0.0 | 0.0±0.0 | 0.0±0.0 | 1.6±1.5 | 2.4±1.9 | 64.5±7.0 | 94.6±1.1 | 91.0±2.2 |
| $T^2=0.20$ | 0.0±0.0 | 0.0±0.0 | 0.0±0.0 | 12.1±6.7 | 43.5±2.4 | 72.2±8.3 | 94.3±0.7 | 92.3±3.1 | 90.0±4.0 |
| $T^2=0.15$ | 0.0±0.0 | 0.0±0.0 | 2.0±1.7 | 42.2±10.5 | 53.1±6.7 | 84.6±8.6 | 92.7±0.9 | 91.7±4.6 | 89.1±5.8 |
| $T^2=0.10$ | 0.0±0.0 | 0.4±0.8 | 20.7±9.4 | 38.9±6.1 | 60.5±2.9 | 85.6±6.4 | 92.9±1.6 | 94.3±1.6 | 90.1±3.7 |
| $T^2=0.05$ | 0.0±0.0 | 20.6±5.7 | 34.8±3.4 | 44.2±2.5 | 66.0±4.3 | 82.3±7.7 | 92.9±2.7 | 93.0±2.0 | 88.7±4.3 |
| $T^2=0.01$ | 57.4±1.4 | 57.7±2.3 | 51.0±1.9 | 53.6±4.6 | 67.4±3.8 | 85.5±4.3 | 92.7±1.5 | 91.6±3.9 | 87.6±3.5 |
| $T^2=0.00$ | 65.4±1.0 | 59.7±1.4 | 51.4±3.4 | 56.4±3.8 | 68.1±3.0 | 87.9±4.0 | 91.1±0.8 | 90.7±3.6 | 86.8±4.6 |

**Table 7.** Accuracy (Mean ± Standard Deviation) table of classification output on test set in target domain (domain adaptation).

| Accuracy (%) | $w_{0,1}=12.0$ $w_2=1.0$ | $w_{0,1}=6.0$ $w_2=1.0$ | $w_{0,1}=3.0$ $w_2=1.0$ | $w_{0,1}=1.5$ $w_2=1.0$ | $w_{0,1}=1.0$ $w_2=1.0$ | $w_{0,1}=1.0$ $w_2=1.5$ | $w_{0,1}=1.0$ $w_2=3.0$ | $w_{0,1}=1.0$ $w_2=6.0$ | $w_{0,1}=1.0$ $w_2=12.0$ |
|---|---|---|---|---|---|---|---|---|---|
| $T^2=0.40$ | 50.0±0.0 | 50.0±0.0 | 50.0±0.0 | 50.0±0.0 | 53.7±4.4 | 75.3±5.4 | 94.3±0.6 | 95.7±0.5 | 92.5±1.4 |
| $T^2=0.20$ | 50.0±0.0 | 50.0±0.0 | 50.2±0.3 | 46.9±26.1 | 50.0±13.0 | 75.1±17.0 | 95.4±0.3 | 94.2±2.0 | 90.7±2.9 |
| $T^2=0.15$ | 50.0±0.0 | 50.0±0.0 | 41.1±16.9 | 43.3±16.2 | 49.9±16.1 | 87.5±10.4 | 94.6±0.5 | 93.3±4.1 | 90.9±3.3 |
| $T^2=0.10$ | 50.0±0.0 | 50.0±0.0 | 51.3±16.5 | 32.5±8.2 | 49.0±4.4 | 91.2±7.4 | 94.8±0.6 | 94.6±1.4 | 91.7±2.1 |
| $T^2=0.05$ | 50.0±0.0 | 57.0±3.2 | 24.8±4.5 | 27.9±3.5 | 65.0±16.5 | 88.3±6.4 | 94.8±0.7 | 94.4±1.9 | 89.8±3.8 |
| $T^2=0.01$ | 39.1±9.3 | 32.0±4.9 | 30.3±3.1 | 40.5±7.5 | 69.2±12.8 | 91.3±3.2 | 94.2±0.5 | 94.5±1.7 | 91.2±1.4 |
| $T^2=0.00$ | 45.0±12.9 | 37.3±2.9 | 31.8±4.6 | 44.0±10.7 | 74.7±8.3 | 93.0±2.3 | 94.2±0.5 | 94.3±1.4 | 89.8±3.1 |

**Table 8.** AUC (Mean ± Standard Deviation) table of classification output on test set in target domain (domain adaptation).

| AUC (%) | $w_{0,1}=12.0$ $w_2=1.0$ | $w_{0,1}=6.0$ $w_2=1.0$ | $w_{0,1}=3.0$ $w_2=1.0$ | $w_{0,1}=1.5$ $w_2=1.0$ | $w_{0,1}=1.0$ $w_2=1.0$ | $w_{0,1}=1.0$ $w_2=1.5$ | $w_{0,1}=1.0$ $w_2=3.0$ | $w_{0,1}=1.0$ $w_2=6.0$ | $w_{0,1}=1.0$ $w_2=12.0$ |
|---|---|---|---|---|---|---|---|---|---|
| $T^2=0.40$ | 49.7±10.1 | 51.9±11.3 | 59.2±9.5 | 47.2±21.9 | 88.7±5.4 | 97.9±1.0 | 97.8±0.5 | 98.9±0.3 | 96.5±1.5 |
| $T^2=0.20$ | 56.4±11.0 | 68.8±9.3 | 67.2±12.9 | 47.5±38.6 | 50.3±20.2 | 77.5±26.8 | 98.9±0.3 | 98.5±0.7 | 94.6±3.6 |
| $T^2=0.15$ | 55.7±8.7 | 55.5±8.6 | 34.9±21.5 | 39.4±25.3 | 47.0±22.3 | 93.6±6.2 | 98.6±0.6 | 97.2±2.7 | 94.6±2.7 |
| $T^2=0.10$ | 44.8±6.2 | 75.5±15.2 | 49.1±23.7 | 19.1±9.2 | 35.5±5.8 | 96.8±3.0 | 98.8±0.5 | 98.7±0.4 | 95.9±2.2 |
| $T^2=0.05$ | 89.7±2.3 | 84.2±13.6 | 8.9±3.6 | 15.2±5.6 | 69.2±21.3 | 95.1±3.2 | 98.9±0.4 | 98.2±0.7 | 94.1±3.3 |
| $T^2=0.01$ | 12.7±6.2 | 20.4±4.4 | 13.2±3.5 | 30.7±9.1 | 73.5±18.3 | 96.8±1.5 | 98.4±1.2 | 98.4±0.6 | 95.9±1.8 |
| $T^2=0.00$ | 43.2±23.5 | 22.3±6.2 | 13.2±5.3 | 34.1±22.9 | 86.1±4.4 | 97.4±1.7 | 98.2±0.6 | 98.3±0.6 | 94.7±1.6 |

**Table 9.** F1-score (Mean ± Standard Deviation) table of classification output on test set in target domain (domain adaptation).

| F1-score (%) | $w_{0,1}=12.0$ $w_2=1.0$ | $w_{0,1}=6.0$ $w_2=1.0$ | $w_{0,1}=3.0$ $w_2=1.0$ | $w_{0,1}=1.5$ $w_2=1.0$ | $w_{0,1}=1.0$ $w_2=1.0$ | $w_{0,1}=1.0$ $w_2=1.5$ | $w_{0,1}=1.0$ $w_2=3.0$ | $w_{0,1}=1.0$ $w_2=6.0$ | $w_{0,1}=1.0$ $w_2=12.0$ |
|---|---|---|---|---|---|---|---|---|---|
| $T^2=0.40$ | 0.0±0.0 | 0.0±0.0 | 0.0±0.0 | 0.2±0.2 | 13.0±13.8 | 66.4±10.1 | 94.1±0.6 | 95.6±0.6 | 91.9±1.7 |
| $T^2=0.20$ | 0.0±0.0 | 0.0±0.0 | 2.1±1.6 | 42.5±23.6 | 62.7±8.2 | 81.0±11.0 | 95.3±0.4 | 93.9±2.3 | 89.8±3.6 |
| $T^2=0.15$ | 0.0±0.0 | 0.0±0.0 | 12.4±10.3 | 53.2±13.5 | 61.5±9.7 | 89.0±7.9 | 94.4±0.4 | 92.8±5.0 | 89.9±4.0 |
| $T^2=0.10$ | 0.0±0.0 | 0.1±0.1 | 33.2±18.0 | 47.2±9.9 | 64.0±3.1 | 91.9±5.7 | 94.6±0.6 | 94.4±1.6 | 91.1±2.5 |
| $T^2=0.05$ | 0.0±0.0 | 23.9±9.6 | 36.8±6.2 | 41.7±3.9 | 73.4±10.2 | 89.3±5.2 | 94.7±0.8 | 94.2±2.2 | 88.5±4.7 |
| $T^2=0.01$ | 6.6±4.0 | 42.9±5.8 | 43.7±3.8 | 53.8±6.8 | 75.1±7.5 | 91.4±2.6 | 94.1±0.6 | 94.2±2.0 | 90.4±1.7 |
| $T^2=0.00$ | 18.7±3.1 | 49.2±2.2 | 45.7±5.4 | 57.4±7.7 | 78.7±5.1 | 93.0±2.1 | 94.2±0.7 | 94.0±1.6 | 88.5±3.9 |

**Table 10.** Accuracy (Mean ± Standard Deviation) table of segmentation output on test set in target domain (domain adaptation).

| Accuracy (%) | $w_{0,1}=12.0$ $w_2=1.0$ | $w_{0,1}=6.0$ $w_2=1.0$ | $w_{0,1}=3.0$ $w_2=1.0$ | $w_{0,1}=1.5$ $w_2=1.0$ | $w_{0,1}=1.0$ $w_2=1.0$ | $w_{0,1}=1.0$ $w_2=1.5$ | $w_{0,1}=1.0$ $w_2=3.0$ | $w_{0,1}=1.0$ $w_2=6.0$ | $w_{0,1}=1.0$ $w_2=12.0$ |
|---|---|---|---|---|---|---|---|---|---|
| $T^2=0.40$ | 50.0±0.0 | 50.0±0.0 | 50.0±0.0 | 50.0±0.0 | 50.2±0.2 | 51.6±0.9 | 79.7±3.5 | 95.6±0.6 | 93.5±0.8 |
| $T^2=0.20$ | 50.0±0.0 | 50.0±0.0 | 50.4±0.5 | 44.0±8.9 | 51.2±8.6 | 70.7±16.5 | 95.7±0.4 | 94.9±1.2 | 92.3±2.5 |
| $T^2=0.15$ | 50.0±0.0 | 50.0±0.0 | 50.2±0.4 | 46.7±13.8 | 51.0±15.3 | 84.0±12.9 | 94.9±0.8 | 94.2±3.1 | 92.8±2.5 |
| $T^2=0.10$ | 50.0±0.0 | 50.1±0.0 | 50.1±4.2 | 29.9±7.4 | 48.4±2.3 | 84.8±10.7 | 94.8±0.8 | 95.3±0.8 | 93.4±1.5 |
| $T^2=0.05$ | 50.0±0.0 | 51.0±8.2 | 24.0±2.9 | 35.0±2.7 | 58.9±10.1 | 80.3±10.6 | 94.9±0.8 | 94.9±1.1 | 91.8±2.3 |
| $T^2=0.01$ | 43.9±1.3 | 45.2±1.1 | 39.6±1.8 | 43.3±3.9 | 62.5±10.1 | 86.0±6.4 | 94.1±0.7 | 95.1±0.9 | 92.7±0.8 |
| $T^2=0.00$ | 49.7±0.3 | 47.3±1.0 | 40.7±2.9 | 45.2±4.8 | 64.6±7.6 | 89.7±4.9 | 93.8±0.7 | 94.6±1.2 | 92.1±2.1 |

**Table 11.** AUC (Mean ± Standard Deviation) table of segmentation output on test set in target domain (domain adaptation).

| AUC (%) | $w_{0,1}=12.0$ $w_2=1.0$ | $w_{0,1}=6.0$ $w_2=1.0$ | $w_{0,1}=3.0$ $w_2=1.0$ | $w_{0,1}=1.5$ $w_2=1.0$ | $w_{0,1}=1.0$ $w_2=1.0$ | $w_{0,1}=1.0$ $w_2=1.5$ | $w_{0,1}=1.0$ $w_2=3.0$ | $w_{0,1}=1.0$ $w_2=6.0$ | $w_{0,1}=1.0$ $w_2=12.0$ |
|---|---|---|---|---|---|---|---|---|---|
| $T^2=0.40$ | 30.6±15.0 | 38.8±19.1 | 46.3±11.7 | 46.7±12.9 | 78.4±10.7 | 89.7±4.0 | 93.8±1.4 | 98.1±0.7 | 95.1±1.6 |
| $T^2=0.20$ | 44.6±20.6 | 32.3±20.3 | 67.6±5.4 | 40.8±26.6 | 52.0±12.5 | 81.1±13.0 | 98.1±0.3 | 97.4±0.7 | 94.4±2.7 |
| $T^2=0.15$ | 48.7±20.6 | 64.0±11.7 | 42.3±14.6 | 45.5±18.6 | 54.1±17.9 | 93.7±5.2 | 97.8±0.4 | 95.9±4.1 | 93.8±3.1 |
| $T^2=0.10$ | 40.1±13.8 | 66.9±17.4 | 39.6±13.1 | 27.3±9.0 | 54.2±5.2 | 95.4±3.1 | 97.6±0.5 | 97.5±0.8 | 94.6±1.9 |
| $T^2=0.05$ | 81.6±5.3 | 53.7±15.3 | 13.7±4.0 | 20.3±5.2 | 71.3±14.0 | 93.9±4.0 | 97.9±0.4 | 97.5±0.9 | 93.5±2.5 |
| $T^2=0.01$ | 33.8±3.9 | 19.9±4.3 | 13.4±5.3 | 27.9±7.0 | 70.6±15.5 | 95.1±0.5 | 97.1±0.6 | 97.3±0.9 | 94.9±1.9 |
| $T^2=0.00$ | 32.7±5.5 | 18.0±4.9 | 14.2±7.7 | 29.1±19.5 | 77.9±6.8 | 95.9±0.5 | 97.2±0.5 | 96.2±2.5 | 92.2±4.1 |

**Table 12.** F1-score (Mean ± Standard Deviation) table of segmentation output on test set in target domain (domain adaptation).

| F1-score (%) | $w_{0,1}=12.0$ $w_2=1.0$ | $w_{0,1}=6.0$ $w_2=1.0$ | $w_{0,1}=3.0$ $w_2=1.0$ | $w_{0,1}=1.5$ $w_2=1.0$ | $w_{0,1}=1.0$ $w_2=1.0$ | $w_{0,1}=1.0$ $w_2=1.5$ | $w_{0,1}=1.0$ $w_2=3.0$ | $w_{0,1}=1.0$ $w_2=6.0$ | $w_{0,1}=1.0$ $w_2=12.0$ |
|---|---|---|---|---|---|---|---|---|---|
| $T^2=0.40$ | 0.0±0.0 | 0.0±0.0 | 0.0±0.0 | 0.0±0.0 | 0.9±0.9 | 6.2±3.4 | 74.6±5.5 | 95.5±0.6 | 93.1±1.0 |
| $T^2=0.20$ | 0.0±0.0 | 0.0±0.0 | 1.9±1.9 | 10.7±3.9 | 45.8±4.3 | 76.8±10.4 | 95.6±0.3 | 94.8±1.3 | 91.8±3.0 |
| $T^2=0.15$ | 0.0±0.0 | 0.0±0.0 | 1.4±0.4 | 43.0±12.1 | 59.9±9.3 | 86.7±9.2 | 94.8±0.8 | 93.9±3.6 | 92.2±2.9 |
| $T^2=0.10$ | 0.0±0.0 | 0.2±0.2 | 17.8±9.3 | 42.0±9.3 | 64.2±1.7 | 87.1±7.5 | 94.7±0.7 | 95.2±0.8 | 93.1±1.8 |
| $T^2=0.05$ | 0.1±0.1 | 23.5±8.3 | 36.1±3.6 | 50.7±2.9 | 70.0±5.4 | 83.7±7.6 | 94.8±0.8 | 94.8±1.3 | 91.2±2.8 |
| $T^2=0.01$ | 61.0±1.3 | 61.5±0.9 | 55.8±1.8 | 58.8±3.7 | 71.7±5.5 | 87.5±5.1 | 94.1±0.5 | 94.9±1.0 | 92.2±0.9 |
| $T^2=0.00$ | 66.4±0.2 | 63.5±1.1 | 57.0±2.9 | 60.6±3.6 | 72.9±4.0 | 90.4±3.9 | 93.9±0.6 | 94.5±1.4 | 91.5±2.5 |

**Table 13.** Accuracy (Mean ± Standard Deviation) table of classification output on validation set in target domain (no domain adaptation).

| Accuracy (%) | $w_{0,1}=12.0$ $w_2=1.0$ | $w_{0,1}=6.0$ $w_2=1.0$ | $w_{0,1}=3.0$ $w_2=1.0$ | $w_{0,1}=1.5$ $w_2=1.0$ | $w_{0,1}=1.0$ $w_2=1.0$ | $w_{0,1}=1.0$ $w_2=1.5$ | $w_{0,1}=1.0$ $w_2=3.0$ | $w_{0,1}=1.0$ $w_2=6.0$ | $w_{0,1}=1.0$ $w_2=12.0$ |
|---|---|---|---|---|---|---|---|---|---|
| $T^2=0.40$ | 50.0±0.0 | 50.0±0.0 | 50.0±0.0 | 50.2±0.4 | 50.1±0.2 | 53.3±1.9 | 74.3±7.8 | 74.7±2.4 | 64.5±3.9 |
| $T^2=0.20$ | 50.0±0.0 | 50.0±0.0 | 50.1±0.2 | 57.7±5.8 | 46.1±6.4 | 54.4±3.0 | 83.9±3.8 | 77.7±1.7 | 61.0±2.5 |
| $T^2=0.15$ | 50.0±0.0 | 50.0±0.0 | 51.2±0.9 | 39.2±6.8 | 39.6±2.0 | 61.8±7.9 | 85.2±2.8 | 74.8±2.9 | 60.1±2.1 |
| $T^2=0.10$ | 50.0±0.0 | 50.0±0.0 | 58.2±1.8 | 34.9±2.3 | 44.4±4.6 | 60.7±7.7 | 85.1±2.0 | 74.0±1.0 | 59.2±2.2 |
| $T^2=0.05$ | 50.0±0.0 | 56.7±1.2 | 35.5±4.0 | 32.4±1.4 | 41.6±1.2 | 67.6±3.2 | 82.6±1.5 | 72.4±2.5 | 61.4±2.6 |
| $T^2=0.01$ | 50.0±4.3 | 57.2±5.1 | 33.1±4.7 | 31.4±3.3 | 38.4±3.8 | 69.3±2.8 | 82.5±5.0 | 72.6±3.3 | 59.9±4.6 |
| $T^2=0.00$ | 45.0±8.8 | 54.8±2.9 | 32.7±3.3 | 32.0±3.5 | 40.9±4.2 | 63.8±8.1 | 83.7±2.6 | 74.8±2.0 | 62.3±3.2 |

**Table 14.** AUC (Mean ± Standard Deviation) table of classification output on validation set in target domain (no domain adaptation).

| AUC (%) | $w_{0,1}=12.0$ $w_2=1.0$ | $w_{0,1}=6.0$ $w_2=1.0$ | $w_{0,1}=3.0$ $w_2=1.0$ | $w_{0,1}=1.5$ $w_2=1.0$ | $w_{0,1}=1.0$ $w_2=1.0$ | $w_{0,1}=1.0$ $w_2=1.5$ | $w_{0,1}=1.0$ $w_2=3.0$ | $w_{0,1}=1.0$ $w_2=6.0$ | $w_{0,1}=1.0$ $w_2=12.0$ |
|---|---|---|---|---|---|---|---|---|---|
| $T^2=0.40$ | 37.5±11.3 | 46.9±12.9 | 34.1±14.6 | 70.1±13.2 | 78.5±7.5 | 80.2±10.4 | 84.9±4.8 | 80.7±6.0 | 74.6±4.3 |
| $T^2=0.20$ | 35.5±16.2 | 37.3±9.9 | 71.0±5.9 | 59.0±10.6 | 58.4±5.2 | 73.0±4.9 | 92.1±2.6 | 84.8±3.2 | 81.0±5.3 |
| $T^2=0.15$ | 47.8±13.5 | 68.4±10.6 | 42.9±5.5 | 46.1±5.9 | 42.9±3.8 | 75.3±5.3 | 91.0±3.5 | 83.5±3.3 | 82.8±5.2 |
| $T^2=0.10$ | 17.6±12.6 | 55.6±17.9 | 62.5±3.2 | 35.8±4.6 | 51.9±5.8 | 72.8±4.5 | 90.5±3.1 | 83.6±5.6 | 78.6±6.6 |
| $T^2=0.05$ | 53.5±12.8 | 74.5±3.7 | 33.2±2.4 | 31.5±3.9 | 46.0±2.4 | 75.8±3.3 | 89.9±2.2 | 86.3±3.2 | 83.7±7.9 |
| $T^2=0.01$ | 45.3±7.3 | 52.6±8.3 | 28.3±2.8 | 32.0±5.3 | 42.8±5.2 | 75.3±3.3 | 90.6±4.3 | 85.1±2.6 | 84.1±2.9 |
| $T^2=0.00$ | 34.9±9.6 | 51.5±5.3 | 28.1±3.9 | 33.8±5.8 | 47.3±6.0 | 72.4±3.7 | 91.7±2.4 | 85.1±2.5 | 80.7±6.3 |

**Table 15.** F1-score (Mean ± Standard Deviation) table of classification output on validation set in target domain (no domain adaptation).

| F1-score (%) | $w_{0,1}=12.0$ $w_2=1.0$ | $w_{0,1}=6.0$ $w_2=1.0$ | $w_{0,1}=3.0$ $w_2=1.0$ | $w_{0,1}=1.5$ $w_2=1.0$ | $w_{0,1}=1.0$ $w_2=1.0$ | $w_{0,1}=1.0$ $w_2=1.5$ | $w_{0,1}=1.0$ $w_2=3.0$ | $w_{0,1}=1.0$ $w_2=6.0$ | $w_{0,1}=1.0$ $w_2=12.0$ |
|---|---|---|---|---|---|---|---|---|---|
| $T^2=0.40$ | 0.0±0.0 | 0.0±0.0 | 0.0±0.0 | 0.8±1.6 | 0.4±0.8 | 12.1±6.4 | 74.6±4.7 | 68.1±3.5 | 44.7±9.8 |
| $T^2=0.20$ | 0.0±0.0 | 0.0±0.0 | 0.4±0.8 | 47.6±8.6 | 56.7±2.8 | 64.5±2.1 | 83.5±3.0 | 72.1±2.5 | 36.0±6.3 |
| $T^2=0.15$ | 0.0±0.0 | 0.0±0.0 | 4.6±3.5 | 49.7±4.1 | 53.7±1.5 | 66.3±4.1 | 84.6±3.0 | 66.9±4.6 | 33.7±5.6 |
| $T^2=0.10$ | 0.0±0.0 | 0.0±0.0 | 35.7±3.3 | 47.5±2.4 | 54.6±2.1 | 65.1±4.5 | 83.6±2.3 | 66.1±1.8 | 30.9±6.0 |
| $T^2=0.05$ | 0.0±0.0 | 24.6±3.8 | 40.2±1.0 | 42.9±1.5 | 52.3±2.8 | 68.8±2.8 | 80.7±2.0 | 61.9±4.8 | 37.1±6.8 |
| $T^2=0.01$ | 26.0±18.9 | 40.6±4.6 | 35.0±3.0 | 41.0±3.6 | 48.4±3.0 | 68.4±3.3 | 81.6±4.3 | 62.4±6.6 | 32.4±12.0 |
| $T^2=0.00$ | 17.2±9.8 | 39.8±3.3 | 34.0±2.7 | 41.5±3.6 | 50.1±3.5 | 65.8±4.1 | 82.1±4.4 | 66.9±3.5 | 39.3±7.9 |

**Table 16.** Accuracy (Mean ± Standard Deviation) table of segmentation output on validation set in target domain (no domain adaptation).

| Accuracy (%) | $w_{0,1}=12.0$ $w_2=1.0$ | $w_{0,1}=6.0$ $w_2=1.0$ | $w_{0,1}=3.0$ $w_2=1.0$ | $w_{0,1}=1.5$ $w_2=1.0$ | $w_{0,1}=1.0$ $w_2=1.0$ | $w_{0,1}=1.0$ $w_2=1.5$ | $w_{0,1}=1.0$ $w_2=3.0$ | $w_{0,1}=1.0$ $w_2=6.0$ | $w_{0,1}=1.0$ $w_2=12.0$ |
|---|---|---|---|---|---|---|---|---|---|
| $T^2=0.40$ | 50.0±0.0 | 50.0±0.0 | 50.0±0.0 | 50.0±0.0 | 50.1±0.2 | 51.2±2.2 | 57.3±7.0 | 52.9±6.7 | 51.6±6.6 |
| $T^2=0.20$ | 50.0±0.0 | 50.0±0.0 | 50.5±1.0 | 53.5±4.2 | 41.8±7.8 | 44.4±3.8 | 61.4±7.2 | 55.2±3.1 | 58.1±7.8 |
| $T^2=0.15$ | 50.0±0.0 | 50.0±0.0 | 50.1±0.2 | 34.3±3.7 | 35.1±1.5 | 46.0±2.1 | 55.4±2.8 | 56.0±4.9 | 64.8±4.0 |
| $T^2=0.10$ | 50.0±0.0 | 50.0±0.0 | 53.5±4.1 | 32.2±2.7 | 39.9±3.5 | 47.7±1.0 | 54.2±2.8 | 55.9±4.8 | 60.1±6.1 |
| $T^2=0.05$ | 50.3±0.2 | 53.9±2.1 | 27.4±1.7 | 35.3±2.3 | 42.5±1.9 | 46.7±1.4 | 58.3±2.8 | 60.8±4.3 | 66.1±7.0 |
| $T^2=0.01$ | 47.9±1.2 | 43.5±3.3 | 32.7±0.7 | 35.2±1.6 | 42.2±1.4 | 47.2±1.3 | 55.7±3.8 | 55.8±4.0 | 60.9±3.9 |
| $T^2=0.00$ | 49.3±0.4 | 40.5±2.8 | 32.7±1.3 | 36.6±2.3 | 42.6±2.3 | 45.7±1.9 | 55.7±4.6 | 55.0±3.5 | 55.3±7.3 |

**Table 17.** AUC (Mean ± Standard Deviation) table of segmentation output on validation set in target domain (no domain adaptation).

| AUC (%) | $w_{0,1}=12.0$ $w_2=1.0$ | $w_{0,1}=6.0$ $w_2=1.0$ | $w_{0,1}=3.0$ $w_2=1.0$ | $w_{0,1}=1.5$ $w_2=1.0$ | $w_{0,1}=1.0$ $w_2=1.0$ | $w_{0,1}=1.0$ $w_2=1.5$ | $w_{0,1}=1.0$ $w_2=3.0$ | $w_{0,1}=1.0$ $w_2=6.0$ | $w_{0,1}=1.0$ $w_2=12.0$ |
|---|---|---|---|---|---|---|---|---|---|
| $T^2=0.40$ | 36.8±9.4 | 36.4±11.3 | 37.3±6.3 | 60.8±20.5 | 59.8±23.2 | 70.7±9.7 | 64.0±4.4 | 69.3±6.7 | 56.4±8.8 |
| $T^2=0.20$ | 38.7±7.4 | 42.7±11.5 | 70.3±15.4 | 47.7±8.5 | 40.9±7.8 | 43.1±6.3 | 85.5±3.2 | 75.9±1.5 | 64.8±6.3 |
| $T^2=0.15$ | 42.7±8.0 | 55.2±17.7 | 66.2±11.5 | 28.9±4.7 | 28.9±2.4 | 52.3±8.2 | 84.4±2.7 | 74.5±3.8 | 69.4±4.3 |
| $T^2=0.10$ | 34.1±7.9 | 51.5±21.7 | 46.4±7.6 | 27.4±7.3 | 36.4±4.4 | 53.2±9.8 | 84.5±2.2 | 74.0±6.9 | 66.0±5.8 |
| $T^2=0.05$ | 62.4±14.8 | 61.4±6.4 | 21.8±2.4 | 23.9±3.7 | 35.0±2.8 | 61.5±5.7 | 86.0±1.1 | 78.3±3.4 | 70.3±8.2 |
| $T^2=0.01$ | 39.0±8.8 | 42.8±7.4 | 20.8±2.0 | 22.4±4.0 | 33.6±4.1 | 64.2±6.6 | 86.8±2.7 | 74.5±1.6 | 67.3±3.5 |
| $T^2=0.00$ | 57.4±8.4 | 40.4±5.1 | 20.9±2.9 | 24.4±4.8 | 38.5±5.9 | 59.1±5.5 | 86.6±2.3 | 74.7±2.9 | 62.2±6.8 |

**Table 18.** F1-score (Mean ± Standard Deviation) table of segmentation output on validation set in target domain (no domain adaptation).

| F1-score (%) | $w_{0,1}=12.0$ $w_2=1.0$ | $w_{0,1}=6.0$ $w_2=1.0$ | $w_{0,1}=3.0$ $w_2=1.0$ | $w_{0,1}=1.5$ $w_2=1.0$ | $w_{0,1}=1.0$ $w_2=1.0$ | $w_{0,1}=1.0$ $w_2=1.5$ | $w_{0,1}=1.0$ $w_2=3.0$ | $w_{0,1}=1.0$ $w_2=6.0$ | $w_{0,1}=1.0$ $w_2=12.0$ |
|---|---|---|---|---|---|---|---|---|---|
| $T^2=0.40$ | 0.0±0.0 | 0.0±0.0 | 0.0±0.0 | 0.0±0.0 | 0.4±0.8 | 4.8±7.6 | 58.8±2.9 | 63.2±4.0 | 57.0±4.2 |
| $T^2=0.20$ | 0.0±0.0 | 0.0±0.0 | 1.9±3.8 | 15.0±13.6 | 43.7±2.7 | 58.6±3.2 | 71.5±3.6 | 66.6±1.7 | 61.8±5.2 |
| $T^2=0.15$ | 0.0±0.0 | 0.0±0.0 | 0.4±0.8 | 37.0±3.3 | 50.1±1.9 | 62.1±1.6 | 68.3±1.2 | 67.1±2.3 | 64.7±1.8 |
| $T^2=0.10$ | 0.0±0.0 | 0.0±0.0 | 21.7±10.7 | 44.9±2.3 | 55.4±3.5 | 64.0±0.3 | 67.5±1.5 | 66.6±2.8 | 62.0±3.8 |
| $T^2=0.05$ | 1.2±1.0 | 20.2±5.3 | 34.6±3.3 | 50.7±2.2 | 59.0±2.0 | 63.4±1.2 | 69.6±1.1 | 69.4±2.7 | 65.9±5.6 |
| $T^2=0.01$ | 64.8±1.1 | 53.9±3.3 | 46.3±0.4 | 50.9±1.8 | 58.6±1.2 | 63.4±0.9 | 68.5±2.1 | 66.4±1.5 | 61.5±1.6 |
| $T^2=0.00$ | 66.1±0.4 | 54.1±2.8 | 46.4±1.6 | 52.4±2.5 | 58.7±2.2 | 62.5±1.6 | 68.5±2.1 | 66.6±0.9 | 59.8±4.6 |

**Table 19.** Accuracy (Mean ± Standard Deviation) table of classification output on test set in target domain (no domain adaptation).

| Accuracy (%) | $w_{0,1}=12.0$ $w_2=1.0$ | $w_{0,1}=6.0$ $w_2=1.0$ | $w_{0,1}=3.0$ $w_2=1.0$ | $w_{0,1}=1.5$ $w_2=1.0$ | $w_{0,1}=1.0$ $w_2=1.0$ | $w_{0,1}=1.0$ $w_2=1.5$ | $w_{0,1}=1.0$ $w_2=3.0$ | $w_{0,1}=1.0$ $w_2=6.0$ | $w_{0,1}=1.0$ $w_2=12.0$ |
|---|---|---|---|---|---|---|---|---|---|
| $T^2=0.40$ | 50.0±0.0 | 50.0±0.0 | 50.0±0.0 | 50.1±0.1 | 50.0±0.0 | 56.1±4.3 | 79.7±8.7 | 80.7±1.7 | 62.1±2.6 |
| $T^2=0.20$ | 50.0±0.0 | 50.0±0.0 | 50.0±0.1 | 58.9±5.6 | 51.6±6.8 | 58.1±3.0 | 87.9±2.5 | 83.1±2.4 | 57.7±1.3 |
| $T^2=0.15$ | 50.0±0.0 | 50.1±0.1 | 50.3±0.3 | 43.8±5.8 | 44.1±2.1 | 69.1±7.9 | 88.6±2.3 | 79.1±2.9 | 57.6±2.6 |
| $T^2=0.10$ | 50.0±0.0 | 50.0±0.0 | 56.5±2.4 | 39.0±2.1 | 50.6±3.6 | 67.6±6.0 | 89.1±1.9 | 79.3±3.7 | 56.9±1.3 |
| $T^2=0.05$ | 50.0±0.0 | 55.6±1.1 | 35.0±4.8 | 35.8±2.1 | 48.4±2.5 | 75.0±2.7 | 88.5±1.3 | 76.7±4.4 | 58.9±2.8 |
| $T^2=0.01$ | 50.5±5.4 | 53.5±5.7 | 31.4±2.7 | 34.5±2.7 | 45.7±3.0 | 78.1±2.7 | 86.7±6.6 | 75.7±3.5 | 57.9±4.6 |
| $T^2=0.00$ | 47.9±8.3 | 53.0±4.0 | 31.1±3.1 | 34.9±2.3 | 47.3±3.8 | 72.8±7.9 | 87.6±3.7 | 79.2±1.5 | 60.2±3.6 |

**Table 20.** AUC (Mean ± Standard Deviation) table of classification output on test set in target domain (no domain adaptation).

| AUC (%) | $w_{0,1}=12.0$ $w_2=1.0$ | $w_{0,1}=6.0$ $w_2=1.0$ | $w_{0,1}=3.0$ $w_2=1.0$ | $w_{0,1}=1.5$ $w_2=1.0$ | $w_{0,1}=1.0$ $w_2=1.0$ | $w_{0,1}=1.0$ $w_2=1.5$ | $w_{0,1}=1.0$ $w_2=3.0$ | $w_{0,1}=1.0$ $w_2=6.0$ | $w_{0,1}=1.0$ $w_2=12.0$ |
|---|---|---|---|---|---|---|---|---|---|
| $T^2=0.40$ | 39.1±13.3 | 47.4±16.5 | 35.5±16.0 | 65.1±17.2 | 75.9±7.3 | 84.0±8.7 | 88.9±3.1 | 86.5±4.4 | 79.6±3.4 |
| $T^2=0.20$ | 37.9±20.3 | 38.6±17.6 | 66.5±7.9 | 60.1±9.4 | 69.6±4.4 | 84.0±2.7 | 94.2±0.6 | 90.4±2.5 | 84.9±5.5 |
| $T^2=0.15$ | 49.8±15.8 | 70.6±11.2 | 37.6±8.6 | 49.9±8.4 | 51.7±5.7 | 86.2±1.4 | 94.0±1.1 | 89.1±3.8 | 87.3±4.5 |
| $T^2=0.10$ | 19.3±11.7 | 52.0±17.8 | 56.7±7.6 | 38.7±3.9 | 61.1±5.0 | 84.5±2.5 | 94.2±1.2 | 89.8±2.9 | 83.7±6.4 |
| $T^2=0.05$ | 55.3±11.5 | 71.9±4.4 | 32.6±4.9 | 33.0±4.4 | 55.2±5.7 | 87.7±1.3 | 93.7±1.0 | 91.3±2.1 | 87.8±5.3 |
| $T^2=0.01$ | 43.3±12.0 | 47.3±7.2 | 27.7±3.8 | 32.3±4.7 | 50.8±4.2 | 86.5±2.5 | 93.7±2.5 | 90.1±2.0 | 87.5±1.8 |
| $T^2=0.00$ | 37.0±10.9 | 48.0±4.2 | 27.6±3.9 | 33.6±3.3 | 55.8±6.6 | 84.5±2.6 | 94.5±1.3 | 89.8±2.5 | 85.6±3.7 |

**Table 21.** F1-score (Mean ± Standard Deviation) table of classification output on test set in target domain (no domain adaptation).

| F1-score (%) | $w_{0,1}=12.0$ $w_2=1.0$ | $w_{0,1}=6.0$ $w_2=1.0$ | $w_{0,1}=3.0$ $w_2=1.0$ | $w_{0,1}=1.5$ $w_2=1.0$ | $w_{0,1}=1.0$ $w_2=1.0$ | $w_{0,1}=1.0$ $w_2=1.5$ | $w_{0,1}=1.0$ $w_2=3.0$ | $w_{0,1}=1.0$ $w_2=6.0$ | $w_{0,1}=1.0$ $w_2=12.0$ |
|---|---|---|---|---|---|---|---|---|---|
| $T^2=0.40$ | 0.0±0.0 | 0.0±0.0 | 0.0±0.0 | 0.3±0.4 | 0.2±0.2 | 20.7±12.9 | 81.0±5.7 | 77.7±1.6 | 39.3±6.9 |
| $T^2=0.20$ | 0.0±0.0 | 0.0±0.0 | 0.2±0.3 | 48.9±8.3 | 62.9±3.2 | 68.2±1.4 | 87.9±1.8 | 80.4±3.3 | 27.2±4.1 |
| $T^2=0.15$ | 0.0±0.0 | 0.3±0.4 | 1.3±1.3 | 54.3±3.3 | 58.5±2.0 | 74.2±4.6 | 88.6±1.9 | 74.2±4.6 | 26.2±7.7 |
| $T^2=0.10$ | 0.0±0.0 | 0.0±0.0 | 32.0±2.3 | 51.9±1.6 | 61.1±1.0 | 72.8±3.5 | 88.8±1.7 | 74.5±5.8 | 24.8±4.0 |
| $T^2=0.05$ | 0.0±0.0 | 21.0±4.5 | 38.0±2.9 | 46.8±1.9 | 59.8±1.5 | 77.6±1.8 | 88.0±1.5 | 69.6±7.4 | 30.2±8.1 |
| $T^2=0.01$ | 29.0±14.3 | 35.7±3.8 | 33.4±4.4 | 44.9±2.3 | 56.7±1.5 | 79.0±2.3 | 87.0±5.0 | 68.2±6.3 | 26.9±12.8 |
| $T^2=0.00$ | 21.8±16.2 | 38.0±4.9 | 33.0±2.3 | 45.0±2.1 | 57.7±3.2 | 75.7±4.8 | 87.7±2.8 | 74.4±2.5 | 33.8±9.6 |

**Table 22**. Accuracy (%, mean ± standard deviation) table of segmentation output on test set in target domain (no domain adaptation).

| Accuracy (%) | $w_{0,1} = 12.0$ $w_2 = 1.0$ | $w_{0,1} = 6.0$ $w_2 = 1.0$ | $w_{0,1} = 3.0$ $w_2 = 1.0$ | $w_{0,1} = 1.5$ $w_2 = 1.0$ | $w_{0,1} = 1.0$ $w_2 = 1.0$ | $w_{0,1} = 1.0$ $w_2 = 1.5$ | $w_{0,1} = 1.0$ $w_2 = 3.0$ | $w_{0,1} = 1.0$ $w_2 = 6.0$ | $w_{0,1} = 1.0$ $w_2 = 12.0$ |
|---|---|---|---|---|---|---|---|---|---|
| $T^2 = 0.40$ | 50.0±0.0 | 50.0±0.0 | 50.0±0.0 | 50.0±0.0 | 50.2±0.2 | 51.2±1.2 | 65.4±7.0 | 58.2±7.0 | 57.4±5.0 |
| $T^2 = 0.20$ | 50.0±0.0 | 50.0±0.0 | 50.7±1.3 | 54.3±3.7 | 46.3±7.1 | 49.2±2.5 | 62.5±7.3 | 58.2±3.0 | 64.2±6.8 |
| $T^2 = 0.15$ | 50.0±0.0 | 50.0±0.0 | 50.0±0.0 | 35.3±1.8 | 41.3±1.5 | 50.8±1.8 | 56.3±2.1 | 58.3±5.4 | 72.1±3.0 |
| $T^2 = 0.10$ | 50.0±0.0 | 50.0±0.0 | 51.0±3.5 | 36.0±3.3 | 46.9±1.7 | 50.0±0.9 | 55.6±1.7 | 59.6±4.0 | 67.4±7.2 |
| $T^2 = 0.05$ | 50.1±0.1 | 53.6±1.4 | 28.1±1.8 | 40.1±1.2 | 47.7±0.6 | 49.8±0.2 | 58.0±3.3 | 63.0±4.0 | 72.3±6.1 |
| $T^2 = 0.01$ | 49.8±0.1 | 42.8±3.8 | 33.1±2.6 | 40.3±1.4 | 46.6±0.7 | 50.4±0.8 | 56.6±3.6 | 59.5±5.8 | 69.7±4.5 |
| $T^2 = 0.00$ | 50.0±0.0 | 41.5±1.0 | 33.4±0.9 | 41.0±1.2 | 47.1±0.9 | 49.5±0.5 | 56.2±3.9 | 58.6±5.4 | 62.3±7.0 |

**Table 23**. AUC (Mean ± Standard Deviation) table of segmentation output on test set in target domain (no domain adaptation).

| AUC (%) | $w_{0,1} = 12.0$ $w_2 = 1.0$ | $w_{0,1} = 6.0$ $w_2 = 1.0$ | $w_{0,1} = 3.0$ $w_2 = 1.0$ | $w_{0,1} = 1.5$ $w_2 = 1.0$ | $w_{0,1} = 1.0$ $w_2 = 1.0$ | $w_{0,1} = 1.0$ $w_2 = 1.5$ | $w_{0,1} = 1.0$ $w_2 = 3.0$ | $w_{0,1} = 1.0$ $w_2 = 6.0$ | $w_{0,1} = 1.0$ $w_2 = 12.0$ |
|---|---|---|---|---|---|---|---|---|---|
| $T^2 = 0.40$ | 30.0±9.8 | 33.2±10.9 | 33.7±7.5 | 57.5±21.8 | 55.2±21.3 | 75.8±7.6 | 73.3±3.7 | 78.1±5.2 | 64.4±5.4 |
| $T^2 = 0.20$ | 33.4±5.6 | 36.8±10.1 | 70.0±17.8 | 49.6±7.6 | 47.1±7.8 | 55.5±7.3 | 91.2±2.0 | 83.5±2.5 | 73.8±7.0 |
| $T^2 = 0.15$ | 38.5±9.4 | 53.5±22.7 | 59.7±9.0 | 30.8±2.8 | 35.6±5.3 | 66.5±10.2 | 90.9±1.7 | 81.9±5.5 | 79.3±3.6 |
| $T^2 = 0.10$ | 33.5±7.4 | 53.8±21.2 | 45.5±8.4 | 29.7±8.3 | 44.3±7.1 | 67.0±9.2 | 91.0±1.1 | 82.6±4.1 | 76.1±7.2 |
| $T^2 = 0.05$ | 70.4±12.0 | 61.8±6.2 | 21.6±2.2 | 24.4±3.4 | 42.7±3.9 | 75.7±3.3 | 91.0±1.0 | 86.0±1.2 | 79.3±5.1 |
| $T^2 = 0.01$ | 43.3±11.7 | 41.5±7.9 | 19.7±2.2 | 21.8±2.7 | 39.1±5.4 | 78.1±4.4 | 91.1±2.6 | 82.6±2.5 | 76.6±3.2 |
| $T^2 = 0.00$ | 55.9±7.6 | 39.2±4.2 | 20.1±2.2 | 22.9±2.2 | 43.6±6.2 | 72.8±4.9 | 91.8±1.9 | 82.5±4.1 | 73.1±4.3 |

**Table 24**. F1-score (Mean ± Standard Deviation) table of segmentation output on test set in target domain (no domain adaptation).

| F1-score (%) | $w_{0,1} = 12.0$ $w_2 = 1.0$ | $w_{0,1} = 6.0$ $w_2 = 1.0$ | $w_{0,1} = 3.0$ $w_2 = 1.0$ | $w_{0,1} = 1.5$ $w_2 = 1.0$ | $w_{0,1} = 1.0$ $w_2 = 1.0$ | $w_{0,1} = 1.0$ $w_2 = 1.5$ | $w_{0,1} = 1.0$ $w_2 = 3.0$ | $w_{0,1} = 1.0$ $w_2 = 6.0$ | $w_{0,1} = 1.0$ $w_2 = 12.0$ |
|---|---|---|---|---|---|---|---|---|---|
| $T^2 = 0.40$ | 0.0±0.0 | 0.0±0.0 | 0.0±0.0 | 0.0±0.0 | 0.8±0.9 | 4.6±4.8 | 67.6±3.6 | 68.0±4.0 | 62.7±3.2 |
| $T^2 = 0.20$ | 0.0±0.0 | 0.0±0.0 | 2.5±4.8 | 16.9±12.8 | 50.3±2.1 | 64.1±1.4 | 72.4±3.7 | 69.3±1.7 | 68.4±4.9 |
| $T^2 = 0.15$ | 0.0±0.0 | 0.0±0.0 | 0.2±0.2 | 36.7±5.2 | 57.0±1.5 | 66.1±0.9 | 69.1±0.9 | 69.4±2.9 | 73.3±2.5 |
| $T^2 = 0.10$ | 0.0±0.0 | 0.2±0.2 | 17.6±7.9 | 48.9±3.0 | 62.4±1.1 | 66.0±0.3 | 68.7±0.9 | 70.0±2.0 | 70.6±4.9 |
| $T^2 = 0.05$ | 0.4±0.4 | 19.7±4.4 | 34.3±3.3 | 55.8±1.4 | 64.1±0.8 | 66.1±0.1 | 69.9±1.6 | 71.6±2.1 | 73.7±4.8 |
| $T^2 = 0.01$ | 66.5±0.1 | 52.6±3.3 | 46.3±4.2 | 56.0±1.6 | 62.9±0.6 | 66.3±0.3 | 69.2±2.0 | 69.8±2.9 | 71.5±2.9 |
| $T^2 = 0.00$ | 66.7±0.0 | 54.0±2.2 | 46.6±1.1 | 56.9±1.4 | 63.2±1.0 | 65.9±0.3 | 69.1±1.8 | 69.7±2.7 | 67.9±3.9 |